\documentclass[useAMS,usenatbib, graphics]{mn2e}
\usepackage{graphicx}
\usepackage{txfonts}
\usepackage{color, supertabular, multicol}

\title[Chemical enrichment in M4]{On the internal pollution mechanisms in 
the globular cluster NGC 6121 (M4): heavy-element abundances and AGB models\thanks{Based on observations taken with ESO telescopes under programmes 072.D-0742, 077.D-0182, 085.D-0205}}
\author[V. D'Orazi et al.]{V. D'Orazi$^{1,2}$\thanks{E-mail:
valentina.dorazi@mq.edu.au}, S.W. Campbell$^{2}$, M. Lugaro$^{2}$, J. C. Lattanzio$^{2}$, M. Pignatari$^{3}$, 
\newauthor{ E. Carretta$^{4}$} \\
$^{1}$Department of Physics and Astronomy, Macquarie University, North Ryde, NSW 2109, Australia\\
$^{2}$Monash Centre for Astrophysics, School of 
Mathematical Sciences, Building 28, Monash University, VIC 3800, Australia\\
$^{3}$Department of Physics, Basel University, Klingelbergstrasse 82, Basel, 4056, Switzerland\\
$^{4}$ INAF Osservatorio Astronomico di Bologna, via Ranzani 1, I 50100, Bologna, Italy \\
}

\begin{document}

\date{Accepted 2013 April 25.  Received 2013 April 24; in original form 2013 January 30}

\pagerange{\pageref{firstpage}--\pageref{lastpage}} \pubyear{2013}

\maketitle

\label{firstpage}

\begin{abstract}

Globular clusters display significant variations in their light-element content, pointing to the existence 
of a second stellar generation formed from the ejecta of an earlier generation.
The nature of these internal polluters is still a matter of debate: the 
two most popular scenarios indicate intermediate-mass AGB stars and fast rotating massive stars.
Abundances determination for some key elements can help distinguish between these competitor candidates.
We present in this paper Y abundances
for a sample of 103 red giant branch stars in NGC 6121. 
Within measurement errors, we find that the [Y/Fe] is constant
in this cluster contrary to a recent suggestion. For a subsample of six stars we also find [Rb/Fe] to be
constant, consistent with previous studies showing no
variation in other $s$-process elements.
We also present a new set of stellar yields for intermediate-mass AGB stellar models of 5 and 6 solar masses, 
including heavy element $s$-process abundances.
The uncertainties on the mass-loss rate, the mixing-length parameter, and the nuclear reaction rates
have a major impact on the stellar abundances. 
Within the IM-AGB pollution scenario, the constant abundance of heavy elements inside the cluster 
requires a marginal s-process efficiency in IM-AGB stars.
Such a constrain could still be satisfied by the present models
assuming a stronger mass-loss rate. The uncertainties mentioned above are limiting 
the predictive power of intermediate-mass AGB models. For these reasons, at the moment
we are not able to clearly rule out their role as main polluters 
of the second population stars in globular clusters.
\end{abstract}


\section{Introduction}\label{sec:introduction}

In the last decades a large number of observational (both photometric and 
spectroscopic) and theoretical studies have been devoted to disentangle 
the complex nature of globular clusters (GCs). With a few outstanding 
exceptions ($\omega$~Centauri, \citealt{jp10}; M22, 
\citealt{marino09}; NGC 1851, \citealt{yg08}; M54, 
\citealt{carretta10a}), GCs are homogeneous in their Fe-peak, heavy 
$\alpha$- (e.g, Ca and Ti) and trans-iron elements produced by the $slow$ and 
the $rapid$ neutron-capture processes (the $s$ and the $r$ processes, e.g., Y, Zr, La, Eu; 
see
Gratton et al. 2004, 2012\nocite{gratton04}\nocite{gratton12} for extensive reviews, and 
\citealt{armosky94}; \citealt{james04}; \citealt{smith08}; 
\citealt{dorazi10}; \citealt{rprocessian11} for the 
analysis focused on the heavy elements in GCs).
On the other hand, GCs exhibit significant star-to-star variations in their 
light-element content, C, N, O, F, Na, Mg, and Al 
(\citealt{cohen99}; \citealt{ivans99}; \citealt{smith05}; 
\citealt{kayser08}; Carretta et al. 2009a,b\nocite{carretta09a}\nocite{carretta09b}). 
Depletions in C, O, and Mg abundances appear together with enhancements in N, 
Na, and Al ({\em the light-element anticorrelations}). These chemical 
features, shared by both unevolved (main-sequence and subgiant, e.g., 
\citealt{gratton01}; \citealt{rc02}) and evolved stars (from the red giant 
branch -RGB- to the horizontal branch, \citealt{marino11}; 
\citealt{gratton11}), indicate that (at least) two different stellar 
generations are currently coexisting in GCs. The first generation (FG) 
stars are characterised by high O (C and Mg) and low Na (N and Al) 
abundances and display the same chemical composition of field stars at 
the corresponding GC metallicity. The second generation (SG) 
stars present instead a N/Na/Al-rich (C/O/Mg-poor) pattern. This indicates 
that they were forged from the ejecta of a fraction of FG stars inside the GC (\citealt{dercole08}; \citealt{carretta10b}), which must have 
experienced H burning at high temperature ($T \gtrsim$30 MK). These stars, 
more massive than those we presently observe  
in GCs ($M\sim$0.8 M$_\odot$), had time to evolve and internal temperatures high 
enough to activate in their interiors the 
CNO, NeNa, and MgAl cycles to enhance N, Na, and Al at the expense of C, 
O, Ne, and Mg (e.g., \citealt{denisenkov89}). The astrophysical site 
where this occurred is still under discussion, with two 
main candidates: intermediate-mass asymptotic giant branch 
stars (IM-AGB, 4 M$_{\odot}$ $\lesssim$ $M$ $\lesssim$ 8 M$_{\odot}$) undergoing 
Hot Bottom Burning (HBB, \citealt{ventura01,ventura02,dercole10}), 
or fast rotating massive 
stars (FRMS, \citealt{decressin07,decressin08}; 
\citealt{krause12}, \citeyear{krause13}). Probing the nature of the internal 
polluters in GCs is fundamental to this field of research 
because of the strong implications related to cluster 
formation and early evolutionary properties.

The Li abundances provide us a powerful tool 
to disentangle between these two different scenarios due to 
the fragile nature of this element. At the high 
temperatures where the 
CNO cycle occurs, it is expected that all Li is destroyed (Li starts burning 
above T$\approx 2.5 \times 10^6$ K). Interestingly, while FRMS destroy Li, 
IM-AGB stars can also 
produce it via the \cite{cf71} mechanism. As a consequence, 
the simultaneous abundance determination of Li and of the elements affected by 
proton captures (e.g., O, Na, Mg, Al, hereafter p-capture elements) may 
supply quite stringent observational constraints to the origin of the 
polluters \citep{dantona08,ventura10}. 
In other words, if there is no Li production within the polluters, we should expect a positive 
correlation between O and Li and a Li-Na 
anticorrelation (\citealt{pasquini05}; \citealt{lind09}; 
\citealt{shen10}). Similarly, ascertaining the abundances of F within a GC 
is critical because the 
production/destruction of F is heavily dependent on the stellar 
mass (see \citealt{smith05}; \citealt{dorazi13} for detailed discussions 
on this topic). Along with Li and p-capture elements, surveys of the 
$s$-process elements in GCs establish an extremely effective 
tracer, because they can deliver further information on the mass range of 
the polluter. 
In the Solar System, three different $s$-process components have been identified to contribute to the abundances above Fe. ($i$) The $weak$ $s$-process component occurs in massive 
stars ($M\gtrsim$8~M$_\odot$), during the convective core He burning 
phase and the subsequent C shell burning (\citealt{raiteri91}; 
\citealt{the07}; \citealt{pignatari10}). 
In the 
Solar System material the $weak$ component accounts for a major 
fraction of the $s$-process isotopes between Fe and Sr (60$<$A$<$90, 
see e.g., \citealt{kappeler11}). 
 ($ii$) The $main$ 
$s$-process component takes place in thermally-pulsing low- and 
intermediate-mass AGB stars and it is responsible, through the 
$^{13}$C($\alpha$,n)$^{16}$O and $^{22}$Ne($\alpha$,n)$^{25}$Mg 
reactions, for the production of the $s$-process elements 
between Sr-Y-Zr and Pb (\citealt{busso99}). 
Finally, ($iii$) AGB stars with low initial metallicity contribute 
to about 50\% of the solar $^{208}$Pb and to most of the solar 
$s$-process Bi, defined as the $strong$ $s$-process component (\citealt{gallino98}).

Neither of the candidate polluters seem able to 
successfully reproduce all the observed features in GCs (e.g., 
\citealt{fenner04}; \citealt{karakas06}), though a growing body of evidence 
seems to converge towards IM-AGB stars as the main polluters 
(\citealt{renzini08}; 
\citealt{dm10}). On the other hand, based on observations of $s$-process 
elements, Villanova \& Geisler (2011, 
hereafter VG11) proposed that in the mildly metal-poor GC M4 
([Fe/H]=$-$1.16, \citealt{harris96} -2010 update) the polluters 
were massive main-sequence 
stars. By analysing FLAMES-Giraffe (R$\sim$20,000) spectra 
for a sample of 23 RGB stars, complemented by another 23 stars from 
\citeauthor{marino08} (2008, hereafter Ma08) observed with UVES 
(R$\sim$45,000), they derived abundances for several key elements, such 
as Li, Na, the sum C+N+O, Y, Zr, and Ba. 
In addition to confirming previous 
results on the light element inhomogeneities, they found that, within 
the observational uncertainties, all the heavy elements do not vary, 
with the exception of Y. They found a difference in the [Y/Fe] 
ratio of $\sim$0.2 dex between FG and SG stars and invoked the $weak$ 
$s$-process from massive stars as responsible for the Y enhancement in 
the second population. 
However, the $weak$ $s$-process is mostly 
made in massive stars at solar-like metallicity.
Given the secondary nature, 
the efficiency of the classic $s$-process in massive stars 
is reduced as metallicity decreases, and its 
contribution to the Y inventory is negligible at the metallicity 
of M4 (e.g., \citealt{prantzos90}; \citealt{raiteri93}).

Furthermore, in the FRMS pollution scenario massive stars can contribute to the
light-element inter-cluster enrichment only when they are on the main sequence. 
They do this via
a slow mechanical wind, delivering to the interstellar medium 
the ashes of the H burning produced via the 
CNO cycle. On the other hand, $s$-process elements are produced  
in the later stages of stellar evolution (during the He core and C 
shell burning phases), and expelled into the interstellar medium during 
the core-collapse supernova explosion (SNII).
Theory suggests that all but perhaps the most massive GCs ($>$ 
10$^6$M$_\odot$, like $\omega$ Cen) cannot retain their SN ejecta (e.g., 
\citealt{rd05}).  In fact, no variations in the $\alpha$-elements are 
detected in M4, as well as in other ``archetypal'' GCs (\citealt{carretta10c}). 
Indeed, the FRMS scenario                                     
predicts no variations in the $s$-process element abundances    
within GCs (\citealt{decressin07}).                           
The only possible contribution via stellar winds to $s$-process 
elements from massive stars is from stars that become type C Wolf-Rayet stars, where the envelope and 
the H-shell layers have been previously lost. 
However, it is only in the advanced stages 
of the evolution of these stars that 
the He shell may show some $s$-process enrichment 
(e.g., \citealt{rauscher02}). 

A further constraint on the possible polluters is given by the observations of Li. 
D'Orazi \& Marino (2010, DM10) determined the Li content in M4 
for a sample of 104 RGB stars, of which 32 were below the RGB bump 
luminosity. They did not detect any Li-Na anticorrelation, with FG and 
SG stars sharing the same Li abundance. The average values are 
A(Li)=1.34$\pm$0.04 and A(Li)=1.38$\pm$0.04, respectively for Na-poor 
and Na-rich stars. This implies that Li should not have been 
destroyed, ruling out FRMS (since they destroy Li) and favouring IM-AGB stars, which can synthesise it. 
Similar results were obtained by 
\cite{mucciarelli11}, who focused on Li abundances from the 
main-sequence up to the RGB and pointed out that there is no variation 
in the Li content across the different stellar generations. 
The same finding is also reported by VG11, who 
found identical Li values for N-rich (A(Li)=0.97$\pm$0.04) and N-poor 
(A(Li)=0.97$\pm$0.03) stars.

Finally, whatever $s$-process donor we consider, 
it would be difficult to     
explain a chemical pattern bearing an enhancement in Y        
without a similar signature in the neighbour elements Sr and Zr.
To shed light on this rather confusing picture
we present Y abundances for a sample of 
103 RGB stars; for a small sub-sample of six stars we were also able to 
gather Rb abundances, whose only study available so far comes from \cite{yong08b}. This paper is organised as follows: in 
Section~\ref{sec:analysis} we describe the sample and the abundance 
analysis procedure; in Section~\ref{sec:results} we illustrate our results  
and compare it to previous 
studies. In Section~\ref{sec:discussion} we provide a summary of 
the current knowledge of the 
abundance pattern in M4, as unveiled from several years of 
high-resolution spectroscopic studies, present 
new IM-AGB models, and discuss their 
strength and weakness in
reproducing the observed features
together with the impact of stellar and nuclear 
uncertainties. In Section~\ref{sec:summary} we present our  
conclusions. 

%__________________________________________________________________
\section{Sample and Abundance analysis}\label{sec:analysis}

Our sample comprises 103 RGB stars whose stellar parameters, metallicity 
and p-capture elements are presented by Ma08\nocite{marino08}, while Li 
abundances are given in DM10\nocite{dm10}. We refer to Ma08 for 
details on target selection, observations, and data reduction procedures. 
Here we just recall that spectra were obtained with FLAMES@VLT (mounted 
at UT2, \citealt{pasquini02}), fiber feeding the UVES high-resolution 
spectrograph (nominal resolution R=47,000). The RED standard setup at 
580nm was employed, providing a spectral coverage from 4760$-$6840~\AA; 
the typical S/N ratios of our targets range from 100-200 per pixel. For 
a small sub-sample of six stars we could also exploit UVES spectra 
acquired by E. Carretta and collaborators (ESO program 085.D-0205); the 
860nm setup results in a spectral coverage from 6600 to 10600~\AA, 
allowing us to include the Rb~{\sc i} line at 7800\AA.

\begin{center}
\begin{figure}
\includegraphics[width=8cm]{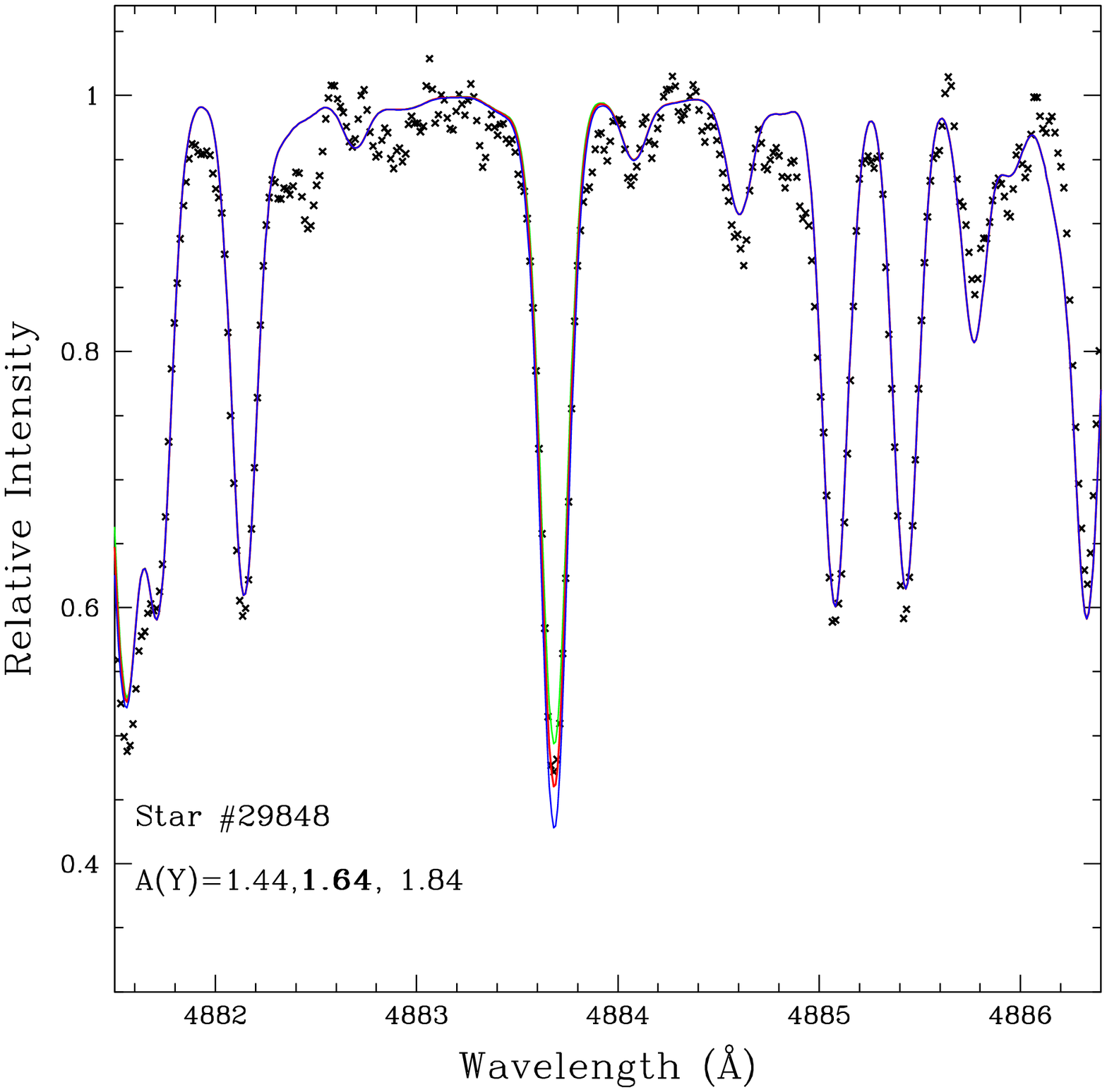}\\
\includegraphics[width=8cm]{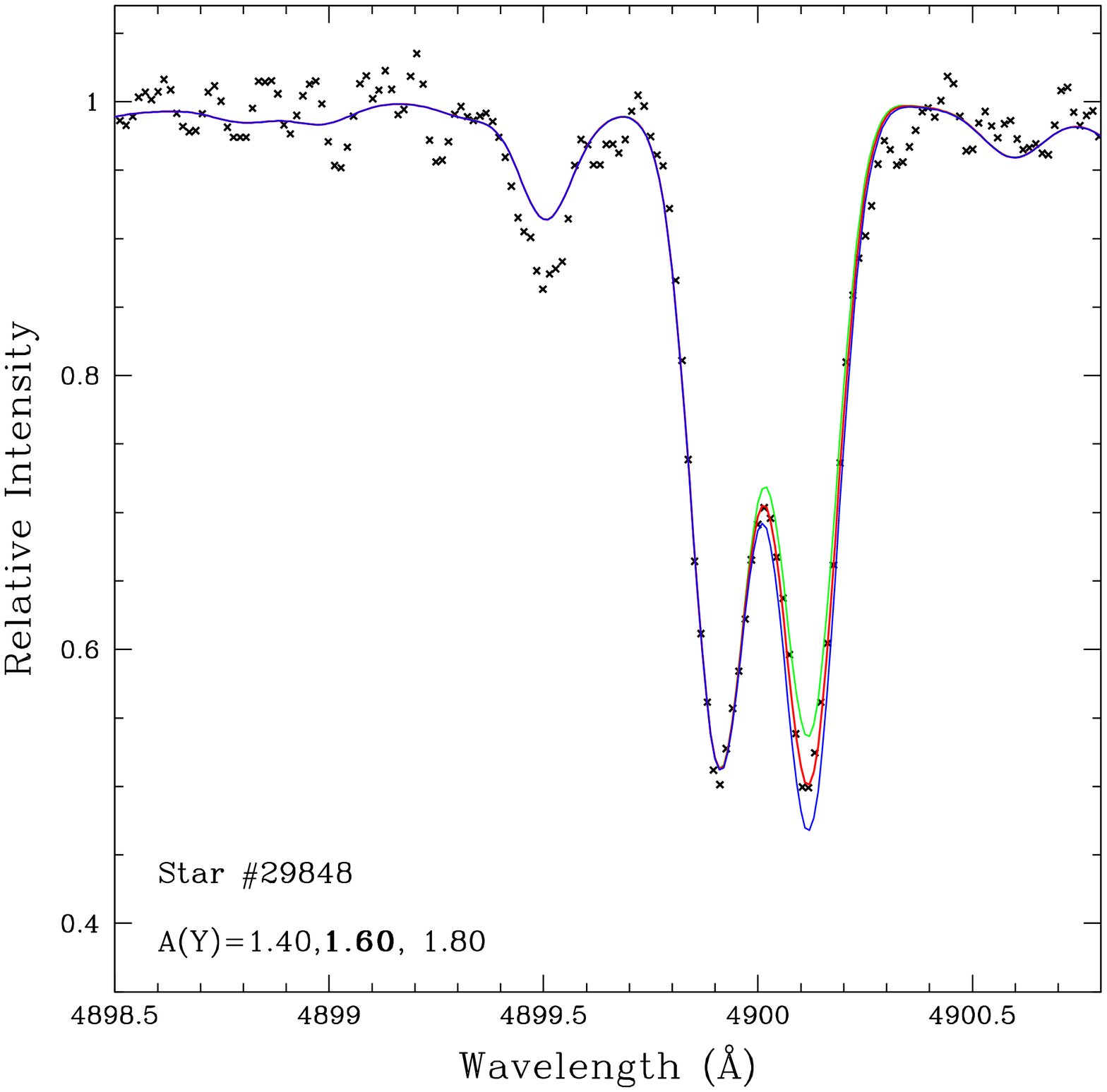}
\caption{Spectral synthesis for the Y~{\sc ii} lines at 4883.68 \AA~and 
4900.12 \AA~for star \#29848.}
\label{f:synth1}
\end{figure}
\end{center}

\begin{center}
\begin{figure*}
\includegraphics[width=13cm]{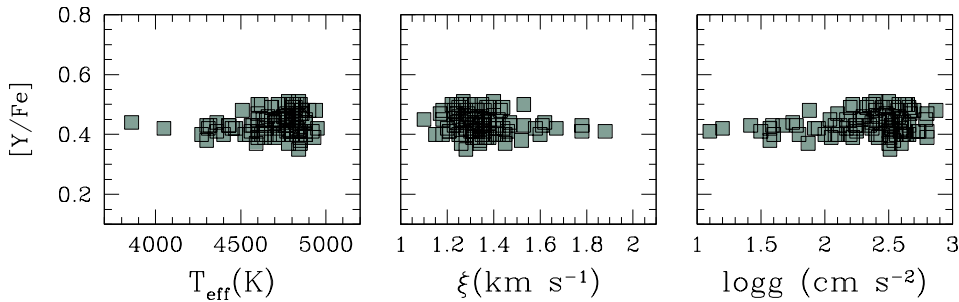}
\caption{Y abundances as a function of stellar parameters ($T_{\rm eff}$, 
microturbulence $\xi$, and log~$g$).}\label{f:param}
\end{figure*}
\end{center}

We derived Y and Rb abundances through the spectral synthesis technique, 
using the code MOOG (\citealt{sneden73}, 2011 version) and the Kurucz 
(\citeyear{kurucz93}) grids of stellar atmospheres, with the 
overshooting option switched off. Although the only stable isotope of Y 
($^{89}$Y) has an odd mass number, the level splitting is 
essentially negligible, because of the small spin and magnetic 
momentum of the Y nucleus. Thus, we adopted a single-line treatment 
focusing on the features at 4883.68~\AA~and 4900.12~\AA. The line atomic 
parameters are listed in Table~\ref{t:list}, where we also report our 
abundances for the Sun and Arcturus ($\alpha$ Bootes). By adopting 
$T_{\rm eff \odot}$=5770~K, log~$g_\odot$=4.44, A(Fe~{\sc 
i})=7.52 and microturbulence $\xi_\odot$=0.9 km s$^{-1}$, we gathered 
A(Y)$_{\odot}$=2.24 from both lines, which is in excellent agreement 
with values from \cite{grevesse96}, \cite{asplund09}, as well as with 
the meteoritic estimates by \cite{lp09}. Regarding Arcturus we 
inferred A(Y)=1.37$\pm$0.04, using as input stellar parameters T$_{\rm 
eff}$=4286 K, log~$g$=1.67, [Fe/H]=$-$0.52, and $\xi$=1.74 km 
s$^{-1}$ (\citealt{ramirez11}). This estimate agrees very well with that 
derived by \cite{smith00}, who focused on lines in the yellow-red 
spectral window and retrieved A(Y)=1.40$\pm$0.15 (see \citealt{dorazi11} 
for more details on the $s$-process elements in Arcturus).  An example 
of spectral syntheses of the Y~{\sc ii} lines for the sample star 
\#29848 is shown in Figure~\ref{f:synth1}, while in Figure~\ref{f:param} 
the resulting abundances for the 103 stars are plotted as a function of 
the stellar parameters.

\begin{center}
\begin{figure}
\includegraphics[width=8cm]{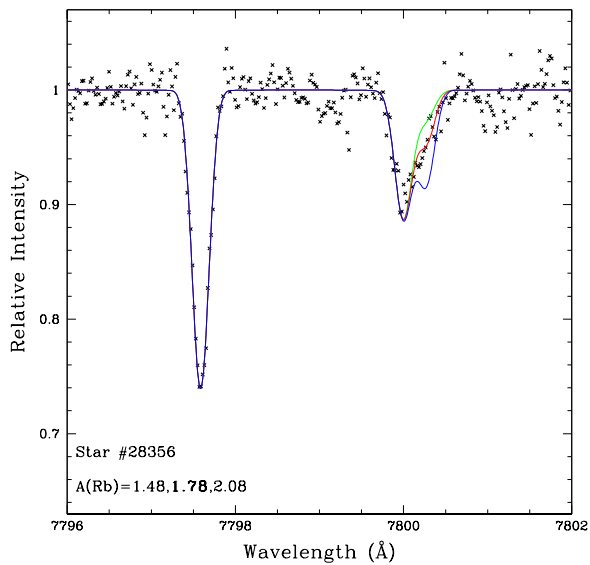}
\caption{Spectral synthesis for the Rb~{\sc i} line at 7800 \AA\ for star \#28356.}
\label{f:synth2}
\end{figure}
\end{center}

\begin{center}
\setcounter{table}{0}
\begin{table}
\caption{ Line list, atomic parameters and abundances for the Sun and Arcturus}\label{t:list}
\begin{tabular}{lccccr}
\hline\hline
Specie              & $\lambda$ & $\chi$ &    log~$gf$    & A(X)$_{\odot}$  & A(X)$_{\alpha{\rm Boo}}$\\
                    &   (\AA)   &  (eV)  &                 &                  &              \\
\hline
Y~{\sc ii}	&    4883.684   &  1.083 &    ~~~0.07      &    2.24	&     1.32              \\
Y~{\sc ii}	&    4900.120   &  1.032 &    $-$0.09	   &	2.24	&     1.42               \\
Rb~{\sc i}	&    7800.268   &  0.000 &    ~~~0.13       &   2.60    &     1.98                \\
\hline\hline
\end{tabular}
\end{table}
\end{center}
The Rb abundance analysis was carried out by synthesising the resonance 
line at 7800.268 \AA, due to the blending with the high excitation 
($\chi$=6.176 eV) Si~{\sc i} line lying on the left wing of that feature 
(at $\lambda$=7799.996 \AA). We took accurate wavelengths and relative 
line strengths of the hyperfine structure components from \cite{ll76}, 
adopting the terrestrial isotopic mixture of $^{85}$Rb/$^{87}$Rb=3 
(e.g., \citealt{tl99}). We included CN features from B. Plez (private 
communication) which significantly improved the continuum fitting; note,
however, that the impact on the Rb syntheses is negligible. Our solar 
analysis results in A(Rb)$_\odot$=2.60, as in \citeauthor{grevesse96} 
(\citeyear{grevesse96}, \citeyear{gs98}), while for Arcturus we derived 
[Rb/H]=$-$0.62 which agrees very well with values published by 
\citeauthor{tl99} (1999, [Rb/H]=$-$0.58), \citeauthor{smith00} (2000, 
[Rb/H]=$-$0.52), and \citeauthor{yong05} (\citeyear{yong05}, 
[Rb/H]=$-$0.55). In Figure~\ref{f:synth2} we show an example of the Rb 
spectral synthesis for star \#28356.

Two kinds of internal errors affect the abundances derived from spectral 
synthesis, those due to the best fit determination and those 
related to the input stellar parameters. In relation to 
the best fit determination, we assumed as a conservative estimate the 
standard deviation from the mean, as given from the two different Y~{\sc 
ii} lines; typical values range from 0.02 to 0.10 dex. The sensitivity 
of the [Y/Fe] ratios to the adopted set of the T$_{\rm 
eff}$, log~$g$, [A/H], and $\xi$ parameters were then assessed in 
the usual way, by  
changing one parameter at the time and 
inspecting the corresponding variation on the resulting abundances. Due 
to the strengths of the Y~{\sc ii} lines, the microturbulence is  
by far the dominant contribution to this error, resulting in 
0.07$-$0.08 dex. Following the error estimates by Ma08\nocite{marino08} 
(i.e., $\Delta$$T_{\rm eff}$=40K, $\Delta$log~$g$=0.12, $\Delta\xi$=0.06 
km s$^{-1}$, and $\Delta$[A/H]=0.05 dex), we found that total 
uncertainty in the [Y/Fe] ratios due to stellar parameters range from 
0.08 to 0.10 dex. Given their independence, we then summed in quadrature 
the two kind of errors, providing the total internal errors as given in 
the last column of Table~\ref{t:results}. Uncertainties in the Rb 
abundances were evaluated following the same approach; because our 
analysis is based on only one line (the Rb~{\sc i} resonance line at 
7947.6\AA\ is too weak to be detected in our sample stars), the errors 
due to the best fitting procedure are 0.10 dex. The total error is given 
in Table~\ref{t:rb_results}.
%____________________________________________________________________

\section{Results}\label{sec:results}
\begin{center}
\begin{table*}
\setcounter{table}{1}
\caption{Stellar parameters, [O/Fe], [Na/Fe] ratios (from Ma08) and 
Y abundances for our 103 sample stars with the related internal errors.}\label{t:results}
\begin{tabular}{lcccccccr}
\hline\hline
 Star   &  $T_{\rm eff}$  & log~$g$         & $\xi$        &  [Fe/H]  &  [O/Fe]&  [Na/Fe] &     [Y/Fe] & Internal\\
        &  (K)   &        & (kms$^{-1}$) &          &        &          &            & error\\
	&        &                &              &          &        &          &            &        \\
\hline	    
  19925 &  4050  &  1.20 &  1.67  &  $-$1.02 &  0.28  &  0.51  &         0.42 &   0.11  \\
  28103 &  3860  &  0.50 &  1.62  &  $-$1.08 &  0.50  &  0.17  &	 0.44 &   0.10	\\
  33414 &  4840  &  2.51 &  1.28  &  $-$1.05 &  0.52  &  0.21  &	 0.35 &   0.13	\\
  35508 &  4780  &  2.48 &  1.18  &  $-$1.05 &  0.25  &  0.38  &	 0.45 &   0.10  \\
  29272 &  4780  &  2.50 &  1.26  &  $-$1.11 &  0.53  &  0.05  &	 0.37 &   0.10	\\
  28797 &  4640  &  2.35 &  1.36  &  $-$1.12 &  0.35  &  0.44  &	 0.43 &   0.09	\\
  29848 &  4780  &  2.52 &  1.24  &  $-$1.05 &  0.54  &  0.09  &	 0.43 &   0.11  \\
   5359 &  4800  &  2.44 &  1.28  &  $-$1.03 &  0.42  &  0.13  &	 0.40 &   0.09	\\
  20766 &  4400  &  1.80 &  1.45  &  $-$1.05 &  0.31  &  0.53  &	 0.40 &   0.09	\\
  21191 &  4270  &  1.60 &  1.60  &  $-$1.06 &  0.34  &  0.51  &	 0.40 &   0.10  \\
  21728 &  4525  &  2.00 &  1.42  &  $-$1.06 &  0.31  &  0.37  &	 0.40 &   0.09	\\
  22089 &  4700  &  2.28 &  1.36  &  $-$1.06 &  0.35  &  0.50  &	 0.45 &   0.10	\\
  24590 &  4850  &  2.66 &  1.35  &  $-$1.07 &  0.50  &  0.30  &	 0.40 &   0.09  \\
  25709 &  4680  &  2.20 &  1.38  &  $-$1.13 &  0.44  &  0.34  &	 0.49 &   0.11	\\
  26471 &  4800  &  2.40 &  1.28  &  $-$1.10 &  0.41  &  0.09  &	 0.48 &   0.09	\\
  26794 &  4800  &  2.45 &  1.44  &  $-$1.17 &  ....  &  0.36  &	 0.49 &   0.10  \\
  27448 &  4310  &  1.57 &  1.58  &  $-$1.12 &  0.51  &  0.11  &	 0.42 &   0.11	\\
  28356 &  4600  &  2.22 &  1.53  &  $-$1.14 &  0.44  &  0.37  &	 0.50 &   0.10	\\
  28707 &  4880  &  2.74 &  1.31  &  $-$1.03 &  ....  &  0.22  &	 0.41 &   0.13  \\
  28847 &  4780  &  2.40 &  1.27  &  $-$1.16 &  0.52  &  0.08  &	 0.51 &   0.11	\\
  28977 &  4680  &  2.33 &  1.40  &  $-$1.14 &  0.39  &  0.40  &	 0.49 &   0.09	\\
  29027 &  4720  &  2.40 &  1.41  &  $-$1.10 &  0.51  &  0.02  &	 0.40 &   0.10	\\
  29065 &  4650  &  2.10 &  1.41  &  $-$1.12 &  0.45  &  0.17  &	 0.40 &   0.09	\\
  29171 &  4880  &  2.64 &  1.26  &  $-$0.99 &  0.22  &  0.41  &	 0.40 &   0.10	\\
  29222 &  4720  &  2.50 &  1.36  &  $-$1.04 &  0.36  &  0.24  &	 0.45 &   0.12	\\
  29282 &  4650  &  2.30 &  1.42  &  $-$1.06 &  0.28  &  0.42  &	 0.45 &   0.09	\\
  29397 &  4600  &  1.50 &  1.78  &  $-$1.12 &  0.24  &  0.46  &	 0.41 &   0.14	\\
  29545 &  4880  &  2.61 &  1.18  &  $-$1.06 &  0.47  & $-$0.02&	 0.40 &   0.12	\\
  29598 &  4840  &  2.50 &  1.40  &  $-$1.06 &  0.39  &  0.40  &	 0.51 &   0.11	\\
  29693	&  4360  &  1.10 &  1.88  &  $-$1.19 &  0.26  &  0.50  &	 0.41 &   0.14	\\
  30209 &  4880  &  2.62 &  1.32  &  $-$0.99 &  0.42  & $-$0.05&	 0.44 &   0.10	\\
  30345 &  4850  &  2.73 &  1.31  &  $-$1.06 &  0.32  &  0.43  &	 0.41 &   0.11	\\
  30450 &  4760  &  2.53 &  1.35  &  $-$1.00 &  0.40  &  0.40  &	 0.46 &   0.11	\\
  30452 &  4830  &  2.56 &  1.25  &  $-$1.00 &  0.40  &  0.06  &	 0.40 &   0.13	\\
  30549 &  4830  &  2.52 &  1.28  &  $-$1.09 &  0.45  &  0.13  &	 0.42 &   0.11	\\
  30598	&  4360  &  1.75 &  1.47  &  $-$1.07 &  0.43  &  0.05  &	 0.44 &   0.09	\\
  30653 &  4660  &  2.30 &  1.25  &  $-$1.06 &  0.47  &  0.15  &	 0.46 &   0.09  \\
  30675 &  4830  &  2.58 &  1.35  &  $-$1.07 &  0.36  &  0.41  &	 0.50 &   0.10	\\
  30711 &  4560  &  2.25 &  1.46  &  $-$1.01 &  0.37  &  0.32  &	 0.41 &   0.09	\\
  30719	&  4810  &  2.65 &  1.24  &  $-$1.19 &  0.44  &  0.42  &	 0.50 &   0.09	\\
  30751 &  4430  &  1.78 &  1.47  &  $-$1.09 &  0.40  &  0.32  &	 0.43 &   0.09	\\
  30924 &  4810  &  2.60 &  1.28  &  $-$1.09 &  0.48  &  0.20  &	 0.45 &   0.09	\\
  30933 &  4800  &  2.63 &  1.30  &  $-$1.07 &  0.44  &  0.44  &	 0.50 &   0.17	\\
  31015 &  4800  &  2.47 &  1.37  &  $-$1.07 &  0.50  &  0.00  &	 0.49 &   0.11	\\
  31306	&  4900  &  2.87 &  1.33  &  $-$1.11 &  0.30  &  0.40  &	 0.48 &   0.13	\\
  31376 &  4800  &  2.59 &  1.36  &  $-$1.00 &  0.33  &  0.43  &	 0.41 &   0.10	\\
  31532 &  4770  &  2.60 &  1.21  &  $-$1.03 &  0.47  &  0.38  &	 0.41 &   0.11	\\
  31665 &  4650  &  2.17 &  1.34  &  $-$1.07 &  0.48  &  0.02  &	 0.42 &   0.10	\\
  31803	&  4850  &  2.60 &  1.34  &  $-$1.13 &  0.38  &  0.25  &	 0.37 &   0.16	\\
  31845 &  4700  &  2.42 &  1.31  &  $-$1.06 &  0.27  &  0.38  &	 0.39 &   0.09	\\
  32055	&  4300  &  1.57 &  1.52  &  $-$1.12 &  0.31  &  0.40  &	 0.38 &   0.10	\\
  32121 &  4840  &  2.58 &  1.33  &  $-$0.93 & ....   &  0.35  &	 0.46 &   0.11		\\
  32151 &  4770  &  2.43 &  1.38  &  $-$1.07 & ....   &  0.39  &	 0.44 &   0.12		\\
  32317 &  4510  &  1.88 &  1.43  &  $-$1.07 &  0.37  &  0.37  &	 0.48 &   0.10	\\
  32347 &  4640  &  2.22 &  1.38  &  $-$1.09 &  0.40  &  0.23  &	 0.39 &   0.09  \\
  32583 &  4850  &  2.54 &  1.34  &  $-$1.08 &  0.46  &  0.07  &	 0.38 &   0.12	\\
  32627 &  4750  &  2.42 &  1.30  &  $-$1.10 &  0.50  & $-$0.03&	 0.46 &   0.11	\\
  32700 &  4560  &  2.12 &  1.36  &  $-$1.02 &  0.49  &  0.06  &	 0.43 &   0.09	\\
  32724 &  4850  &  2.73 &  1.30  &  $-$1.03 &  0.51  &  0.11  &	 0.41 &   0.10	\\
  32782	&  4880  &  2.60 &  1.24  &  $-$1.05 &  0.29  &  0.46  &	 0.44 &   0.10	\\
\end{tabular}
\end{table*}
\setcounter{table}{1}
\begin{table*}
\caption{Continued}\label{t:results}
\begin{tabular}{lcccccccr}
\hline\hline
 Star   &  $T_{\rm eff}$  & log~$g$         & $\xi$        &  [Fe/H]  &  [O/Fe]&  [Na/Fe] &     [Y/Fe] & Internal\\
        &  (K)   &         & (kms$^{-1}$) &          &        &          &            & error\\
	&        &                &              &          &        &          &            &        \\
\hline	    
  32871 &  4770  &  2.48 &  1.24  &  $-$1.01 &  0.40  &  0.06  &	 0.47 &   0.17	\\
  32874 &  4600  &  2.04 &  1.38  &  $-$1.14 &  0.41  &  0.11  &	 0.45 &   0.09	\\
  32933 &  4430  &  1.42 &  1.78  &  $-$1.13 &  0.48  &  0.02  &	 0.43 &   0.13	\\
  32968 &  4630  &  2.17 &  1.30  &  $-$1.13 &  0.41  &  0.24  &	 0.45 &   0.09	\\
  32988	&  4850  &  2.63 &  1.22  &  $-$1.09 &  0.51  &  0.02  &	 0.50 &   0.10	\\
  33069 &  4940  &  3.05 &  1.36  &  $-$0.92 &  0.39  &  0.23  &	 0.48 &   0.09	\\
  33195 &  4620  &  2.38 &  1.26  &  $-$1.03 &  0.48  &  0.09  &	 0.50 &   0.11	\\
  33617 &  4720  &  2.35 &  1.26  &  $-$1.09 &  0.29  &  0.37  &	 0.50 &   0.09	\\
  33629	&  4930  &  2.80 &  1.33  &  $-$0.98 &  0.40  &  0.31  &	 0.39 &   0.09	\\
  33683 &  4800  &  2.57 &  1.18  &  $-$1.05 &  0.42  &  0.00  &	 0.48 &   0.09	\\
  33788 &  4700  &  2.37 &  1.33  &  $-$1.02 &  0.30  &  0.37  &	 0.44 &   0.09	\\
  33900 &  4770  &  2.48 &  1.27  &  $-$1.06 &  0.30  &  0.47  &	 0.45 &   0.09	\\
  33946	&  4800  &  2.62 &  1.15  &  $-$1.03 &  0.23  &  0.32  &	 0.40 &   0.10  \\
  34006	&  4320  &  1.67 &  1.61  &  $-$1.06 &  0.25  &  0.44  &	 0.43 &   0.11	\\
  34130 &  4550  &  2.08 &  1.40  &  $-$1.09 &  0.33  &  0.43  &	 0.40 &   0.09	\\
  34167	&  4950  &  2.60 &  1.40  &  $-$1.10 &  ....  &  0.13  &	 0.42 &   0.13		\\
  34240 &  4470  &  1.95 &  1.41  &  $-$1.10 &  0.47  &  0.08  &	 0.42 &   0.10	\\
  34502	&  4860  &  2.70 &  1.33  &  $-$1.08 &  0.35  &  0.34  &	 0.48 &   0.09	\\
  34726 &  4600  &  2.24 &  1.35  &  $-$1.01 &  0.37  &  0.42  &	 0.43 &   0.09	\\
  35022	&  4850  &  2.51 &  1.36  &  $-$1.08 &  0.30  &  0.41  &	 0.39 &   0.10	\\
  35061	&  4860  &  2.67 &  1.23  &  $-$0.99 &  0.50  &  0.06  &	 0.48 &   0.10	\\
  35455 &  4600  &  2.10 &  1.29  &  $-$1.06 &  0.45  &  0.01  &	 0.47 &   0.09	\\
  35487 &  4850  &  2.67 &  1.24  &  $-$1.00 &  0.36  &  0.30  &	 0.47 &   0.10	\\
  35571	&  4880  &  2.79 &  1.10  &  $-$0.99 &  0.45  & $-$0.02&	 0.45 &   0.10	\\
  35627	&  4830  &  2.37 &  1.20  &  $-$1.08 &  ....  &  0.04  &	 0.42 &   0.09		\\
  35688 &  4720  &  2.25 &  1.33  &  $-$1.11 &  0.21  &  0.40  &	 0.42 &   0.09	\\
  35774 &  4450  &  1.92 &  1.44  &  $-$1.10 &  0.43  &  0.24  &	 0.42 &   0.09	\\
  36215	&  4300  &  1.59 &  1.53  &  $-$1.11 &  0.48  &  0.18  &	 0.43 &   0.09	\\
  36356 &  4820  &  2.66 &  1.26  &  $-$1.05 &  0.30  &  0.26  &	 0.49 &   0.09	\\
  36929 &  4820  &  2.55 &  1.28  &  $-$1.03 &  0.34  &  0.45  &	 0.44 &   0.10	\\
  36942	&  4800  &  2.66 &  1.23  &  $-$0.98 &  0.55  &  0.04  &	 0.43 &   0.13  \\
  37215	&  4790  &  2.50 &  1.21  &  $-$1.11 &  0.45  &  0.25  &	 0.41 &   0.13	\\
  38075 &  4800  &  2.54 &  1.25  &  $-$1.07 &  ....  &  0.42  &	 0.49 &   0.13		\\
  38383 &  4590  &  1.87 &  1.45  &  $-$1.10 &  0.39  &  0.11  &	 0.37 &   0.13	\\
  38896 &  4760  &  2.53 &  1.31  &  $-$1.02 &  0.31  &  0.43  &	 0.43 &   0.09	\\
  42490 &  4570  &  2.08 &  1.41  &  $-$1.07 &  0.29  &  0.38  &	 0.43 &   0.09	\\
  42620 &  4600  &  2.05 &  1.37  &  $-$1.09 &  0.22  &  0.48  &	 0.39 &   0.11	\\
  43370	&  4920  &  2.80 &  1.32  &  $-$1.05 &  0.20  &  0.34  &	 0.41 &   0.11	\\
  44243	&  4860  &  2.80 &  1.17  &  $-$1.04 &  ....  &  0.41  &	 0.47 &   0.10		\\
  44595 &  4750  &  2.40 &  1.40  &  $-$1.07 &  0.20  &  0.44  &	 0.40 &   0.09	\\
  44616 &  4620  &  2.20 &  1.44  &  $-$1.04 &  0.28  &  0.44  &	 0.40 &   0.10	\\
  45163 &  4770  &  2.40 &  1.26  &  $-$1.10 &  ....  &  0.46  &	 0.45 &   0.12		\\
  45895 &  4720  &  2.25 &  1.34  &  $-$1.04 &  0.43  &  0.10  &	 0.43 &   0.09	\\
	&        &	 &	  &	     &	      &	       &	      & 	\\
\hline\hline
\end{tabular}	
\end{table*}
\end{center}	

\begin{center}
\begin{figure*}
\includegraphics[width=13cm]{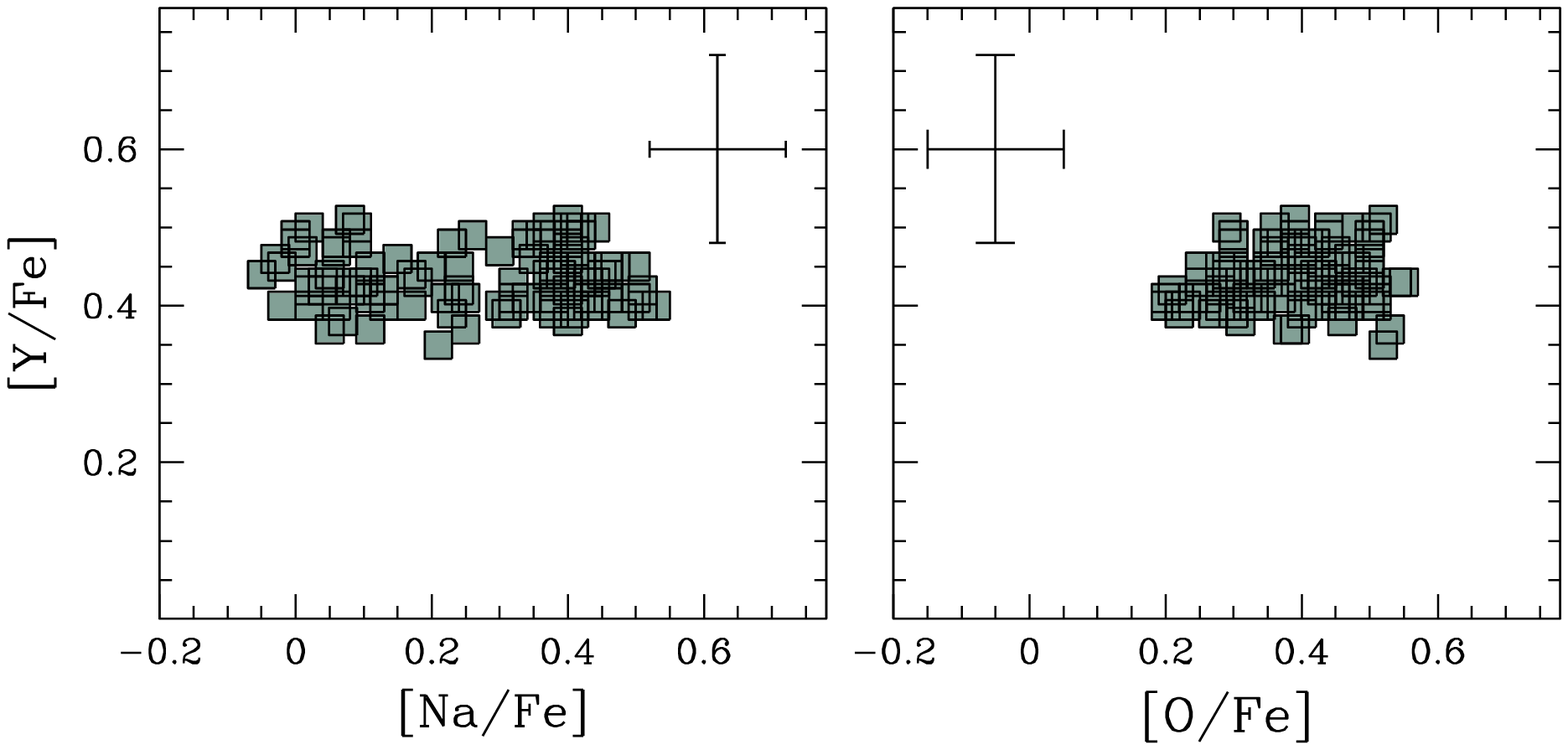}
\caption{[Y/Fe] (this study) vs [Na/Fe] and [O/Fe] (from Ma08) for our sample stars.}\label{f:YNAO}
\end{figure*}
\end{center}

Our results are presented in Table~\ref{t:results}, where we report 
the stellar parameters, the Na and O abundances from Ma08 and our [Y/Fe] 
ratios along with the corresponding internal uncertainties (computed as 
detailed in Section \ref{sec:analysis}). Considering the whole sample of 
103 stars, we inferred a mean value of [Y/Fe]=0.436$\pm$0.004 
(rms=0.038). Our main result is that the scatter is significantly smaller 
than the internal uncertainties, indicating that {\em there is no Y 
variation among our sample stars.} If we divide FG and SG stars, 
according to their [Na/Fe] ratios as done by Ma08 and VG11 (with FG 
stars defined by having [Na/Fe]$<$0.23 dex), we obtain
[Y/Fe]=0.434$\pm$0.006 
(rms=0.034) and [Y/Fe]=0.437$\pm$0.005 (rms=0.038) 
for FG and SG stars, respectively. Our finding implies that in 
M4, while we observe a conspicuous variation in p-capture elements, the 
two different stellar generations share the same Y abundances; this is 
shown in Figure~\ref{f:YNAO}, where the [Y/Fe] ratios are plotted as a 
function of [Na/Fe] and [O/Fe] from Ma08. The variations among different stars 
of 0.58 and 0.35 dex for Na and O are not accompanied by a 
change in the Y abundances (see discussion in 
Section~\ref{sec:discussion}).

\begin{center}
\begin{figure}
\includegraphics[width=8cm]{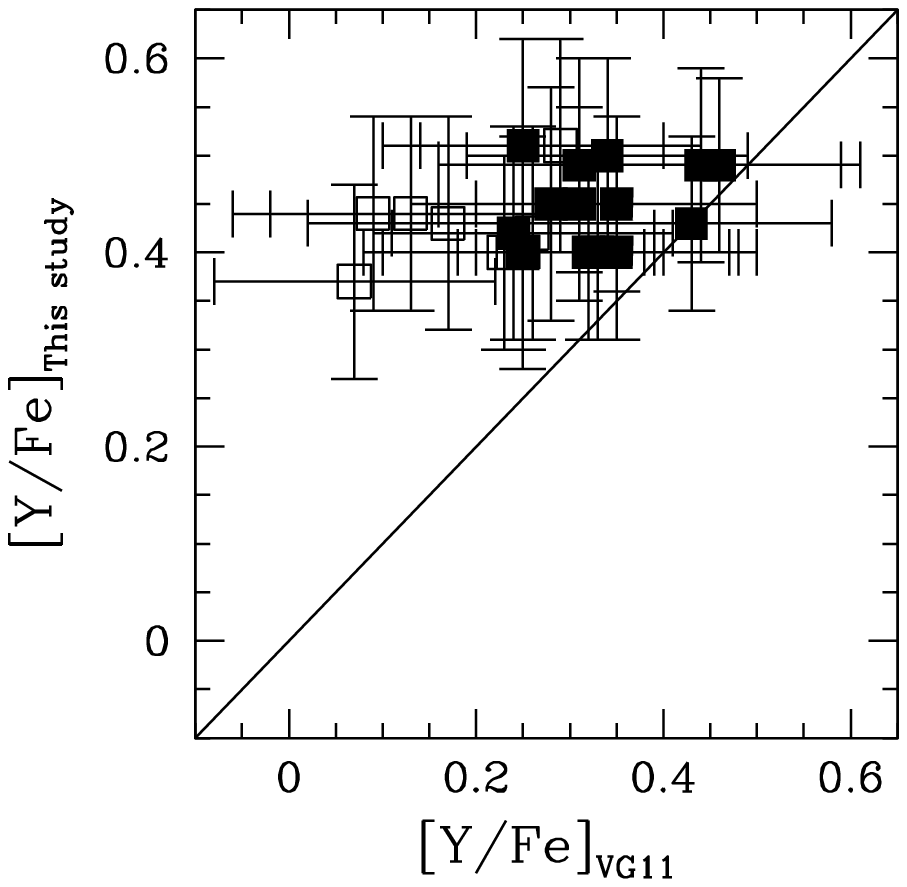}
\caption{Comparison between our Y abundances and those by VG11 for FG (empty squares) and SG stars 
(filled squares).}
\label{f:compVG11}
\end{figure}
\end{center}

The evidence we collected from our measurements on the constancy of Y 
within M4 is in disagreement with results obtained by VG11. 
The 23 RGB stars observed with UVES by VG11 are also included 
in our sample. 
In Figure~\ref{f:compVG11} we compare the Y abundances from the two studies: 
there is a discrepancy between the 
two estimates, which is larger for the Na-poor (i.e., FG) stars.
Focusing only on their UVES sample, VG11 found 
[Y/Fe]=0.20$\pm$0.03 and [Y/Fe]=0.34$\pm$0.02 for FG and SG 
stars, respectively.
\begin{center}
\begin{figure*}
\includegraphics[width=13cm]{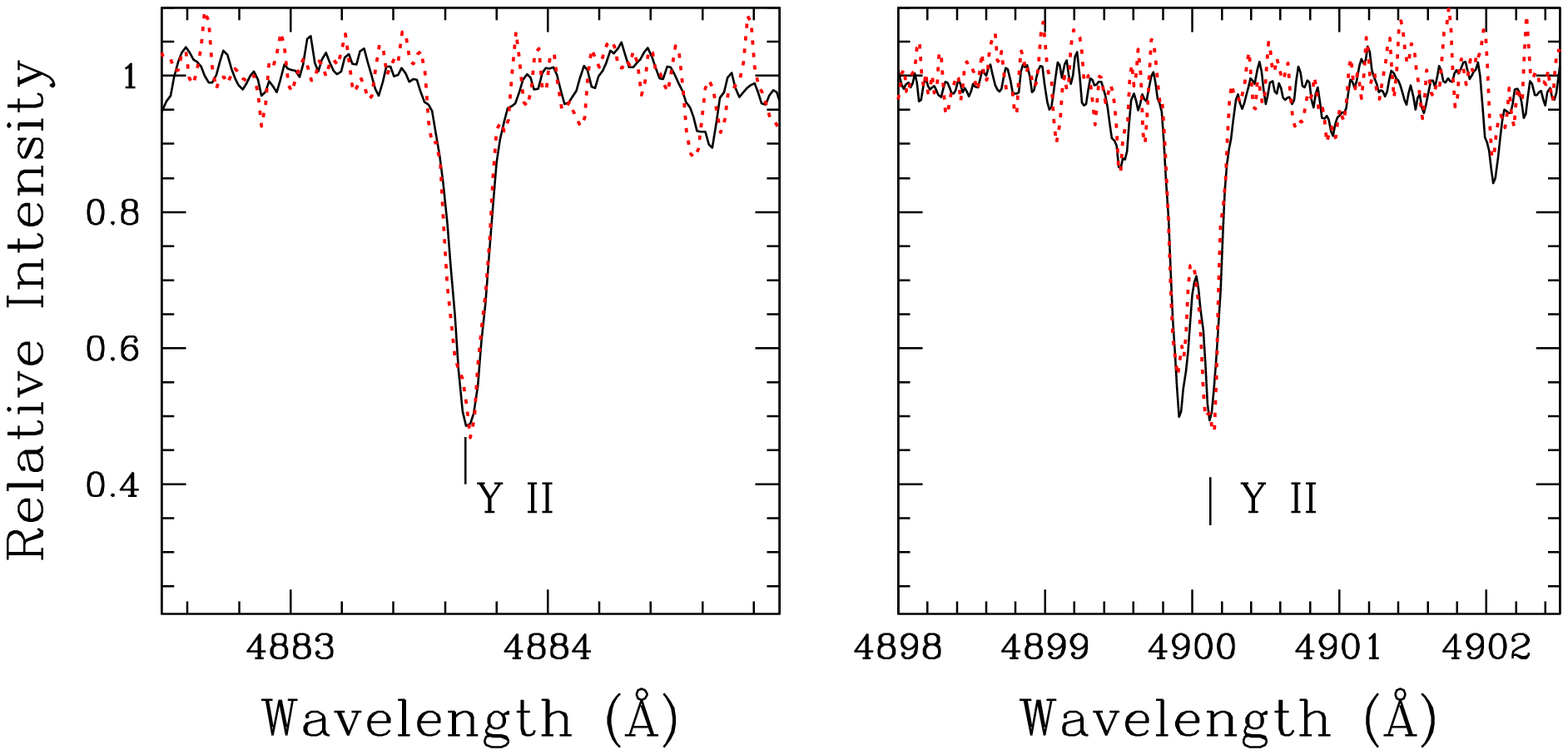}
\caption{Comparison of the two Y~{\sc ii} lines for two stars of almost identical stellar parameters (\#29848, solid line; \#26794, dotted line). }\label{f:compYlines}
\end{figure*}
\end{center}

There are no obvious explanations for the origin of this difference since we used the same spectra, the same Y~{\sc ii}
line at 4900\AA, model atmospheres, 
and stellar parameters (all from Ma08), as well as the same 
abundance code, employing the spectral synthesis technique. 
In Figure~\ref{f:compYlines} we show the direct comparison of the spectra for two 
stars with very similar atmospheric parameters, and for which VG11 have 
derived a difference in the [Y/Fe] ratios of $\Delta$[Y/Fe]=0.27, being 
[Y/Fe]=0.17 for \#29848 and [Y/Fe]=0.44 for \#26794. 
As can be 
seen from the figure, the Y~{\sc ii} lines are of the same 
strength in both stars and we indeed obtained [Y/Fe]=0.43$\pm$0.11 
and [Y/Fe]=0.49$\pm$0.10, respectively. 
Furthermore, the low [Y/Fe] ratios derived by VG11 for FG stars 
(ranging from 0.07 to 0.29) does not fit the global picture of the 
neutron-capture element abundances in M4. In fact, it is well assessed 
from several studies that M4 is characterised by an intrinsically high 
$s$-process element content, compared to other GCs (like e.g., its {\it 
twin} M5, see \citealt{bw92}; \citealt{ivans99}, \citeyear{ivans01}; 
Yong et al. 2008a,b\nocite{yong08a}\nocite{yong08b}; \nocite{marino08}Ma08; 
\citealt{smith08}; \citealt{dorazi10}).

\begin{center}
\begin{figure}
\includegraphics[width=8cm]{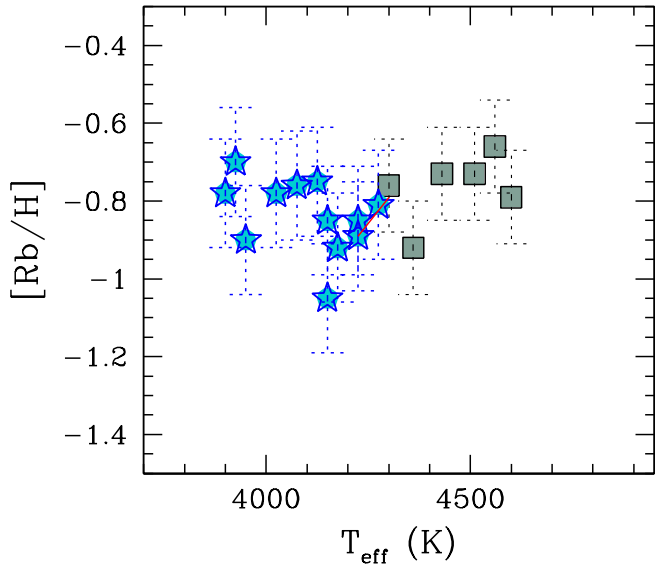}
\caption{[Rb/H] ratios as a function of temperature for our stars (squares)
and for the 12 giants from Yong et al. (2008a, starred symbols)}\label{f:Rb}
\end{figure}
\end{center}

On the other hand, five of the stars in our sample 
are in common with  
\cite{yong08b}. The average [Y/Fe] in \cite{yong08b}
and in our study is 0.69$\pm$0.02 and 0.42$\pm$0.01 dex, respectively, with
a difference of 0.27 dex. Such a difference is mainly due 
to a [Fe/H] offset of 0.18$\pm$0.02 dex, whereas                 
the average [Y/H] values differ only by 0.08$\pm$0.02 dex, within the observational uncertainties 
in the adopted stellar parameters. 
For instance, $T_{\rm eff}$ values of Yong and collaborators are on average cooler than Ma08
and log~$g$ are systematically lower by $-$0.43$\pm$0.07.
\begin{center}
\setcounter{table}{2}
\begin{table}
\caption{Rb abundances for a sub-sample of six stars (see text)}\label{t:rb_results}
\begin{tabular}{lcr}
\hline\hline
 Star & [Rb/Fe]  & Internal \\
      &          &  error    \\
\hline
      &          &                \\
28356 &  0.35	 &	0.10	  \\
29693 &  0.27	 &	0.12	  \\
30711 &  0.35	 &	0.10	  \\
30751 &  0.36	 &	0.11	  \\
32317 &  0.34	 &	0.10	  \\
36215 &  0.32	 &	0.12	  \\
\hline\hline
\end{tabular}
\end{table}
\end{center}
For 6 stars we derived an average value of [Rb/Fe]=0.34$\pm$0.01. 
Although based on a quite limited sample,                     
the Rb abundance is constant within M4, confirming previous   
findings by \cite{yong08a}. One of our star is in common 
with that study: \#36215 (L3624 in Yong et al.'s sample) for which
we derived [Rb/H]=$-$0.79 to be compared with [Rb/H]=$-$0.89 by   
Yong et al.
In general there is a good agreement between the
two estimates: our mean value is [Rb/H]=$-$0.77$\pm$0.04 (rms=0.09),
to be compared to [Rb/H]=$-$0.84$\pm$0.03 (rms=0.09)  (see 
Figure~\ref{f:Rb}).
%
%

%_________________________________________________________________________________________
%
%
\section{Discussion}\label{sec:discussion}

M4 is perhaps the most extensively studied GC. Mainly thanks to its 
close distance (R$_{\rm Sun}$=2.2 kpc, \citealt{harris96}), a wealth of 
spectroscopic studies have been accomplished, aimed at 
deriving information on its abundance pattern for species from Li up to 
Pb. This cluster represents an excellent example of what we deem as an 
{\em archetypal} GC, because it presents substantial 
changes only in p-capture elements, while heavy elements do not show any 
variations beyond what is expected from observational errors. A further 
confirmation of the ``standard'' nature of M4 can also be found 
in the recent survey by Carretta et al. (2012a,b), who 
performed a careful analysis of the Al content in three GCs. In 
NGC~6752 and 47 Tuc (NGC~104) at least three episodes of star formation 
are required to account for the chemical pattern. In M4 the  
classical picture of a FG plus a SG
formed from the ejecta (diluted to a certain extent) of FG stars is able 
to reproduce the observed trends. It is worth mentioning, in this context, that the cluster is however characterised 
by an intrinsically high level of $s$-process elements, when compared to other clusters like M5 (which has a very similar metallicity, [Fe/H]$\approx-$1.2 dex; see e.g., \citealt{ivans99}, \citeyear{ivans01}). Several studies have been devoted to assess this very peculiar heavy element pattern, 
indicating that the proto-cloud from which M4 formed must have been enriched by an higher concentration of $s$-process elements (\citealt{yong08b}).

In the following Section~\ref{sec:observations} we briefly                       
overview the current knowledge of                             
the chemical composition of this cluster.                     
In Section~\ref{sec:models} we present a new set of intermediate-mass 
AGB models and discuss their comparison with the observational data to
challenge the IM-AGB pollution scenario.

\subsection{The chemical abundance pattern of M4}\label{sec:observations}
\subsubsection*{Light elements}\label{sec:light}
\begin{center}
\begin{figure*}
\includegraphics[width=13cm]{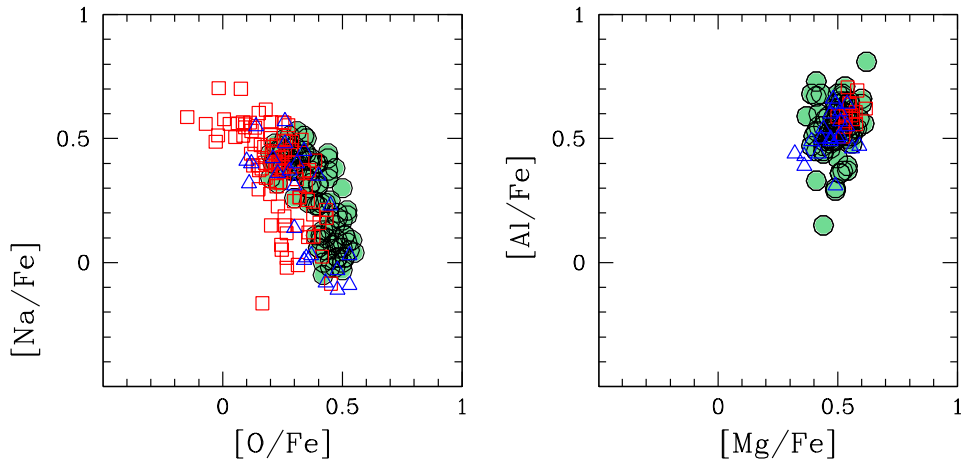}
\caption{Na-O anticorrelation (left panel) and 
[Al/Fe] as a function of [Mg/Fe] (right panel). 
Data are from Ma08 (filled circles); Carretta et al. (2009a,b, empty squares); 
VG11 (empty triangles). The relatively small offsets between the different studies can provide a conservative estimate of the measurement
uncertainties.}\label{f:nao}
\end{figure*}
\end{center}
The most recent abundance determinations for the key elements Na, O, Mg, 
Al were presented by Ma08\nocite{marino08}, Carretta et al. (2009a,b) and VG11\nocite{vg11}. Ma08 analysed a 
sample of 104 giants from high-resolution UVES spectra and detected a well 
defined Na-O anticorrelation (left panel of 
Figure~\ref{f:nao}), with [O/Fe] ranging from 0.20 to 0.55 
($\Delta$=0.35 dex) and [Na/Fe] ranging from $-$0.05 to 
0.53 dex ($\Delta$=0.58 dex). Very similar results were obtained by 
VG11, who confirmed 0.10$<$[O/Fe]$<$0.53 and 
$-$0.11$<$[Na/Fe]$<$0.57, while slightly larger variations in the Na-O 
plane were measured by \cite{carretta09a}. Based on the  
intermediate-resolution FLAMES-Giraffe (R$\sim$20000) spectra of 88 
giants, and a different temperature scale based on photometry, they found the [O/Fe] ratio varying from $-$0.15 to 0.45 
($\Delta$=0.60 dex) and [Na/Fe] from $-$0.16 to 0.70 dex 
(i.e., $\Delta$=0.86 dex). While all the three works detected the 
presence of a very clear Na-O anticorrelation, there is no consensus on 
Al. Ma08 found a variation in [Al/Fe] 
(at $\sim$0.4 dex), positively correlated with 
Na, as also previously detected by \cite{ivans99}, but no significant (anti)correlation with the 
Mg abundances. At variance with that study, both \cite{carretta09b} and 
VG11 did not detect any significant variation in the Al content of the 
cluster (right panel of Figure~\ref{f:nao}). Carretta et al.  
presented Mg and Al abundances for a sample of 14 RGB stars from UVES 
spectra and obtained constant values for [Mg/Fe]=0.55$\pm$0.01 and 
[Al/Fe]=0.60$\pm$0.01; analogously VG11 found no hint of Al variation, 
with FG and SG stars showing [Al/Fe]=0.51$\pm$0.04 and 
[Al/Fe]=0.53$\pm$0.02, respectively. 
%These authors widely discussed that 
%the discrepancy between their results and those presented by Ma08 
%might be related to the presence of unidentified molecular features 
%(possibily CN) blended with the Al lines (the doublet at 6696/6698\AA), 
%whose strength depends critically on the effective temperature, with the 
%cooler stars more heavily affected (see VG11 for details). 
On the other hand,  
Carretta et al. (2012b)  presented Al abundances for 83 RGB stars 
derived from the strong doublet at 8772-8773\AA, properly taking into 
account the CN features populating that spectral region. They detected a 
relatively small variation in Al, positively correlated with Na, 
and confirmed the constancy of the Mg abundances 
([Mg/Fe]=+0.541$\pm$0.005 dex). In general, beyond the debated 
question of the existence of an Al variation, all these studies 
agree that the cluster does not show the Mg-Al anticorrelation.

As previously mentioned in Section~\ref{sec:introduction}, Li abundances 
in M4 were derived from three different studies: DM10, 
\cite{mucciarelli11}, and VG11. Despite the presence of systematic 
offsets between the different studies (Li is critically dependent on the 
adopted effective temperature scale), the main result 
is that there is no Li-Na anticorrelation and no Li-O 
correlation (which would be expected in the case of Li destruction within 
the polluters), since both FG and SG stars show the same Li content. 
While DM10 
and VG11 analysed the Li content in RGB stars (both below and above the 
RGB bump luminosity), \cite{mucciarelli11} gathered Li abundances 
for a sample of dwarf members, and report A(Li)=2.30$\pm$0.02 dex 
(rms=0.10), consistent with the {\it Spite plateau} defined by halo 
dwarfs. The fact that FG and SG stars exhibit the same Li  
calls for yields from the polluters to be at roughly the Spite plateau 
level and led DM10 to speculate that relatively low-mass polluters 
($\approx$4 M$_{\odot}$) were at work in M4. Further support to this 
suggestion comes from the modest Na-O anticorrelation (with 
the absence of extremely O-poor stars, as compared to other GCs, such 
as NGC~2808, \citealt{bragaglia10}) and from the lack of the Mg-Al 
anti-correlations (see, however, the next Section for 
comparison between models and observations).

Concerning the C and N abundances, the cluster is known to exhibit a very 
well defined bimodal distribution in the CN band strengths, as first 
discovered by \cite{norris81}. \cite{ivans99} found that variations in 
CN are not random but they are positively correlated with Na and Al (with the 
CN-strong stars showing also higher abundances of Na and Al) and 
anticorrelated with O. A similar result was later presented by Ma08 
who, adopting the CN strength index S(3839) as given by Smith \& 
Briley (2005), showed the (anti)correlation of CN index with (O)Na (see 
Figure~10 of that paper). The sum of C+N+O has been found to be 
constant, within the uncertainties, by \cite{ivans99} and VG11, who 
obtained A(C+N+O)=8.24$\pm$0.03 and A(C+N+O)=8.16$\pm$0.02, 
respectively. We note, in passing, that the constancy of the C+N+O sum in GCs 
is currently debated. A most striking case is NGC~1851, 
for which Yong (2011) detected an huge spread of about 1.6 
dex\footnote{${\rm 
http://www.ucolick.org/kraftfest/talks/yong\_kraftfest.pdf}$}, while 
\cite{villanova10} claimed a constant value.

Finally, the He content has been measured for six blue HB stars from 
\cite{villanova12}, who found these stars all enhanced in He by 
0.04 dex with respect to the primordial He abundance (i.e., 
Y=0.24-0.25). The same stars are also Na-rich (O-poor), as expected 
according to the multiple population scenario (we refer the reader to 
\citealt{gratton10} and \citealt{marino11} for a discussion on the relationship between 
light-element variations, HB morphology, and the {\em second-parameter} 
problem).

\subsubsection*{Heavy elements}

\begin{center}
\begin{figure}
\includegraphics[width=8cm]{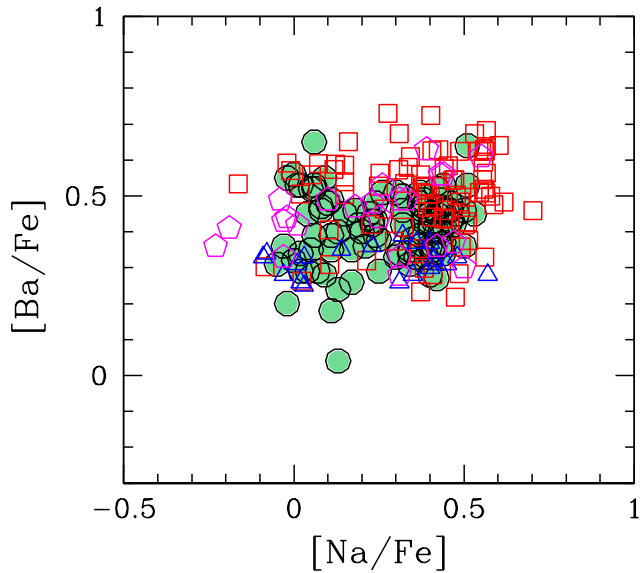}
\caption{[Ba/Fe] vs [Na/Fe]. Symbols are as for Figure~\ref{f:nao} for Ma08 and VG11, while 
empty squares and pentagons are for D'Orazi et al. (2010) and Ivans et al. (1999), respectively.}\label{f:BaFe}
\end{figure}
\end{center}

Yong et al. (2008a,b) published a comprehensive abundance study for a 
sample of twelve RGB stars in M4, deriving abundances for 27 elements, 
from Si to Hf. These authors concluded that there is no change in any of 
the $n$-capture elements under scrutiny; regarding Y they found a mean 
value of [Y/Fe]=0.69$\pm$0.02 (rms=0.05). The FG stars 
(i.e., [Na/Fe]$<$0.30) exhibit [Y/Fe]=0.67$\pm$0.03 (rms=0.05) while 
Na-rich, SG stars have [Y/Fe]=0.70$\pm$0.02(rms=0.06). The Pearson's 
correlation coefficient between [Y/Fe] and [Na/Fe] is 
very small, showing that there is no statistically significant relationship between 
the two. Similarly, none of the other heavy elements 
show any significant variation, with 
[Sr/Fe]=0.73$\pm$0.04, [Zr/Fe]=0.48$\pm$0.03 (from Zr~{\sc ii} lines), 
[La/Fe]=0.48$\pm$0.03, [Eu/Fe]=0.40$\pm$0.03, and [Pb/Fe]=0.30$\pm$0.02 dex.

Barium abundances were published by \cite{ivans99}, 
Ma08\nocite{marino08}, \cite{dorazi10}, and VG11\nocite{vg11} who found 
[Ba/Fe]=0.60$\pm$0.02, [Ba/Fe]=0.41$\pm$0.01, [Ba/Fe]=0.50$\pm$0.01, and 
[Ba/Fe]=0.32$\pm$0.02, respectively. We show in Figure~\ref{f:BaFe} the 
[Ba/Fe] ratio as function of the [Na/Fe] ratio, 
reporting measurements from these 
previous studies. Although the Ba abundances are characterised 
by relatively large uncertainties (due to the saturated behaviour of the 
Ba~{\sc ii} lines), there is no obvious relationship with the Na 
abundances (nor with other p-capture elements).

A key element in this context is Rb ($Z$=37), whose abundance in the 
solar system material is $\sim$50\% $s$-process 
(\citealt{simmerer04}). Theory predicts that AGB stars with 
$M\gtrsim$5M$_\odot$ overproduce Rb with respect to solar and the 
other nearby $s$-process elements (e.g., Y, Zr). This is the result of 
neutron densities reaching up to 10$^{14}$ n/cm$^3$ and favouring the 
operation of the branching points on the $s$-process path at $^{85}$Kr 
and $^{86}$Rb (\citealt{abia01,vanraai12}). 
As previously mentioned in Section~\ref{sec:results}, \cite{yong08b} 
obtained that the Rb content does not show any cosmic star-to-star scatter
 (i.e., [Rb/Fe]=0.39$\pm$0.02, rms=0.07). For FG and SG stars, 
the average ratios are [Rb/Fe]=0.37$\pm$0.02 (rms=0.04) and 
[Rb/Fe]=0.40$\pm$0.02 (rms=0.07), respectively. We confirm this previous 
finding and find a mean abundance of [Rb/Fe]=0.34$\pm$0.01.

In summary, all the previous surveys on heavy-element element 
abundances in M4, with the exception of Y by VG11, agree that 
these elements do not show any intrinsic inhomogeneity, within the observational 
uncertainties. This demands that the polluters must 
account for changes in p-capture elements without 
affecting the n-capture elements (see Section~\ref{sec:models}).

Finally,  
since it is not the goal of this paper to discuss the 
intrinsic enrichment of heavy elements in M4
and to address the origin of the first generations of stars 
responsible for such a peculiar signature,
we refer to previous papers available in the literature on this topic 
(e.g., Yong et al. 2008a,b; \citealt{karakas10}).
We simply take these abundances of light and heavy elements 
as initial composition and conditions to study the chemical 
enrichment inside M4 (see section 4.2).

%
%___________________________________________________________________________________________

\subsection{Intermediate-mass AGB models}\label{sec:models}

\subsubsection{The stellar models}

\begin{table*}
\caption{List of our grid of models. Final $M_{core}$ is the mass of the
  remnant white dwarf.}\label{t:models}
    \begin{tabular}{lclclcc}
        \hline
        LABEL & Initial Mass & AGB Mass-loss & $\alpha_{MLT}$ & Ne Rates & No. TPs & Final $M_{core}$\\ 
        \hline
        STD5                                  & 5.0 &\cite{vw93}&1.75 & Standard & 79  & 0.91  \\ 
        MLT5                        & 5.0 &\cite{vw93} & 2.20 & '' & 73 & 0.91\\ 
        BLK5                         & 5.0 & \cite{bloecker95} & 1.75 & '' & 39 & 0.91 \\ 
        M+B5 & 5.0 &\cite{bloecker95} & 2.20 & '' & 29 & 0.91 \\ 
        \hline
        STD6                                  & 6.0 & \cite{vw93} & 1.75 & '' & 76 & 1.00 \\ 
        MLT6                        & 6.0 & \cite{vw93} & 2.20 & '' & 55 & 0.99 \\ 
        BLK6                         & 6.0 & \cite{bloecker95} & 1.75 & '' &  47 & 0.99 \\ 
        M+B6 & 6.0 &\cite{bloecker95} & 2.20 & '' & 32 &0.99 \\
        MBR & 6.0 &\cite{bloecker95} & 2.20 & $^{22}$Ne(p,$\gamma$)$^{23}$Na high$^a$ & 32 &0.99 \\
          &  &  &  & \& $^{23}$Na(p,$\alpha$)$^{24}$Mg low$^a$ & &  \\ \hline
        \hline
$^a$As given by \citet{iliadis10}. 
    \end{tabular}
\end{table*}

We have computed a grid of stellar models to investigate the scenario where
the material that went into forming the second generation stars is assumed
to come from the ejecta of a population of IM-AGB stars.  To compute the
stellar structure models we used the Monash version of the Monash-Mount
Stromlo evolutionary code (MONSTAR, see eg.  \citealt{Wood1981, 
lattanzio86, frost96}) including recent updates as described by
\citet{campbell08}. Low-temperature opacities have been further updated to
those calculated by \cite{lederer09}, which are variable in C and N as
needed to follow the pollution of the stellar surface via the third
dredge-up (TDU) episodes in AGB stars \citep{marigo02}. Instantaneous
mixing was used in convective zones and the convective boundaries were
defined using a search for ``convective neutrality'' \citep{frost96}. No
further overshoot was applied beyond this boundary.  Initial composition
for the models was based on the observed FG abundances reported by 
VG11 (see their Table 2). This composition is alpha-enhanced and has
sub-solar C/Fe.  Since the initial [O/Fe] $= +0.4$, the metallicity Z is
roughly doubled from that given by [Fe/H] only and the present models have initial
Z $= 0.002$.  In our standard models mass-loss was included using the
empirical formula of \citet{reimers75} during the RGB phase (with
$\eta = 0.4$) and the empirical formula of \citet{vw93} during the AGB
phase. The stellar evolution was followed from the zero-age main sequence
to the end of the thermally-pulsing AGB phase. In some cases the
evolution could not be followed past the instability caused by the peak in
Fe opacity in the AGB envelope (see \citealt{lau12} for a discussion) and
final envelope masses between 0.1 $\textrm{~M}_{\odot}$ and 1.2
$\textrm{~M}_{\odot}$ remained at the end of the calculations. In these
cases when calculating the yields we assume that the surface abundances in
this remaining material stay the same as those in the last computed model.

Models of AGB stars are well known to contain many significant
uncertainties.  These include, but are not limited to: (1) the convection
theory, in these models we used the mixing length theory
(MLT), (2) the mass-loss rates, and (3) the nuclear
reaction rates.  Our grid of models explores
the effects of these three uncertainties. We calculated two `standard'
(STD) models of initial masses 5 M$_\odot$ and 6 M$_\odot$ using the input
physics described above and setting the mixing length parameter
$\alpha_{MLT} = l_{MLT}/H_P$ (where $l_{MLT}$ is the mixing length and
$H_{P}$ is the pressure scale-height) to 1.75, as calibrated using a model
of the Sun.  As a first test, we increased $\alpha_{MLT}$ to
2.20. Variations of $\alpha_{MLT}$ to this extent are warranted since
$\alpha_{MLT}$ probably changes for different evolutionary phases
\citep{lydon93,lebzelter07} and metallicities \citep{chieffi95,palmieri02,
  ferraro06,tramp11} and it is expected to change for AGB stars
\citep{sackmann91,lebzelter07}.  \citet{ventura05a} found that the
$\alpha_{MLT}$ value corresponding to employing the full spectrum of
turbulence formalism (FST, \citealt{canuto91}) in massive AGB stars is $\sim$ 2
(although this does not reproduce all the features of FST
models). \citet{mcsaveney07} used values of $\alpha_{MLT}$ up to 2.6 to
match observations of the effective temperature of the envelopes of massive
AGB stars in the Magellanic Clouds.  As a second test, we used the mass-loss formula of
\citet{bloecker95}, with $\eta = 0.02$ during the AGB phase.  This
mass-loss prescription gives higher mass-loss rates than the formula of
\citet{vw93}, effectively reducing the lifetime of the star. The star
consequently suffers fewer thermal pulses and therefore fewer TDU episodes,
which follow each TP (see also \citealt{ventura05b}).  As a third test, we changed \emph{both} the
mass-loss rate to \cite{bloecker95} and $\alpha_{MLT}$ to 2.2.

We also performed detailed nucleosynthesis calculations using a
post-processing code, the Monash Stellar Nucleosynthesis code (MONSOON;
\citealt{Cannon1993, Lattanzio1996, Lugaro2004}), that reads as input the
background thermal structure from the stellar evolution calculations. The
code solves a network of 2,336 nuclear reactions involving 320 nuclear
species from n and p up to Bi. The bulk of the reaction rates are from the
JINA reaclib database as of May 2009. We tested that there are no major
changes when using the database as of May 2012. 
During the post-processing we made one final test to investigate the
effects of changing the p-capture rates that affect the production of
$^{23}$Na, as detailed in the last line of Table \ref{t:models} (model
MBR, see also \citealt{ventura05b}; 2006).  We summarise the 9 computed models and their parameters in Table
\ref{t:models}.

\subsubsection{Model results}

In the last two columns of Table \ref{t:models} we list the total
number of thermal pulses and the final core mass for each model.  The
models show little difference in final core mass because the core grows
very slowly due to the presence of deep TDU episodes, which 
removes the core growth due to H burning (i.e., 
$\lambda_{TDU} \sim 1$). The mass of the core is not affected by higher
mass-loss rates since it essentially reaches its final value after $\sim$ 10 TPs. On the
other hand, the different models present large differences in the number of
TPs.  Increasing $\alpha_{MLT}$ caused a minor reduction in number of TPs
in the 5 M$_\odot$ model while in the 6 M$_\odot$ model the number of TPs
was reduced by $\sim 25\%$.  This is a result of hotter HBB and 
higher luminosities driving an
enhanced mass loss \citep{ventura05a}. In the models with increased
mass-loss rates, due to the use of the \cite{bloecker95} formula, obviously
the effect is stronger.  In both the 5 M$_\odot$ and 6 M$_\odot$ models the
number of TPs as compared to the standard models is significantly lower,
reducing from $79 \rightarrow 39$ TPs and $76\rightarrow 47$ TPs,
respectively. As expected, the combination of a higher $\alpha_{MLT}$ and a
higher mass-loss rate leads to an even further reduction of the number of
TPs (Table \ref{t:models}).

\begin{figure}
\centering
\includegraphics[width=\columnwidth]{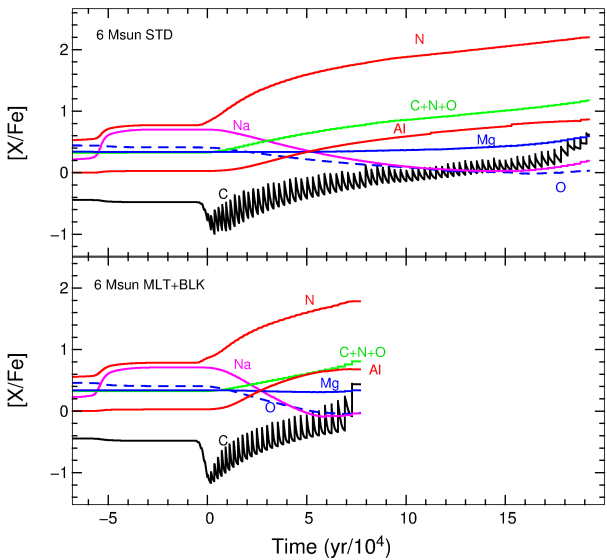}
\caption{The evolution of the surface C, N, O, Na, Mg, Al, and the sum of
  the C+N+O abundances in the standard 6 M$_\odot$ model (STD6, upper
  panel) and in the model computed using the Bloecker mass-loss rate
  formula and $\alpha_{MLT} = 2.2$ (model M+B6, lower panel). Time is
  offset so $t=0$ coincides with the start of the AGB for both models. We
  note that the Al in the plot is mostly $^{26}$Al which, due to its relatively short
  half-life, was assumed to decay completely to $^{26}$Mg and was thus
  added to the yields of Mg presented in Figs.~\ref{f:yldmgal5} and 
\ref{f:yldmgal6}.}\label{f:srf1}
\end{figure}

\begin{figure}
\centering
\includegraphics[width=\columnwidth]{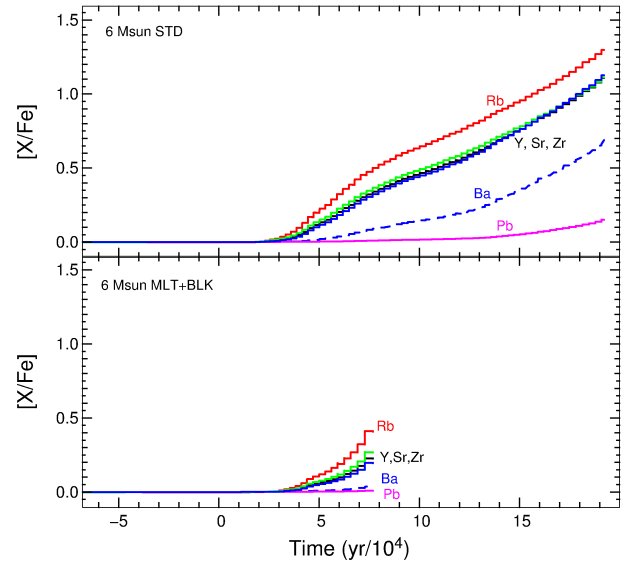}
\caption{Same as Figure \ref{f:srf1} for selected $s$-process elements
  belonging to the first (Rb, Sr, Y, and Zr) second (Ba) and third (Pb)
  $s$-process peaks.}\label{f:srf2}
\end{figure}

In Figures \ref{f:srf1} and \ref{f:srf2} we compare the evolution of the
surface abundances along the AGB in the standard 6 M$_\odot$ model (STD6)
and in the 6 M$_\odot$ model computed using Bloecker mass-loss and
$\alpha_{MLT} = 2.2$ (M+B6). Using the formula of \cite{bloecker95} and
$\alpha_{MLT} = 2.2$ severely truncates the AGB phase, resulting in fewer
TDU episodes and less material carried into the envelope. The abundances of species arising from the TDU are lower,
e.g., the sum of C+N+O (Figure  \ref{f:srf1}) and the $s$-process elements
(Figure \ref{f:srf2}). The sum of C+N+O is determined by the amount of C
produced by He burning during the TPs, dredged-up to the stellar surface,
and partially converted to N by HBB. Thus, the truncation of the AGB phase
naturally leads to a significantly smaller increase in the sum of C+N+O.
The higher value of $\alpha_{MLT}$ also has the effect of decreasing the
number of TDU episodes and, in addition, it has the important consequence
of altering the thermal structure of the convective envelope. In particular, 
the temperature at the base of the envelope increases. This allows more
efficient HBB, which modifies the nucleosynthetic yields of all the species
affected by proton captures, for example, the surface O abundance reduces
faster (Figure \ref{f:srf1}).
 
The $s$-process elements in our IM-AGB star models are produced in the
convective regions associated with the TPs.  The neutron source during the
TPs is the $^{22}$Ne($\alpha$,n)$^{25}$Mg reaction. The $^{22}$Ne is
synthesised in the TPs by conversion of all the $^{14}$N, produced by the
H-burning shell during the interpulse period, via two $\alpha$
captures. 
An important fraction of the $^{22}$Ne abundance is of primary origin
since it is formed from $^{14}$N which is produced via proton captures on 
the $^{12}$C left behind
by He burning and carried to the envelope by the TDU.
%
%The $^{22}$Ne abundance also has an important component
%independent of the initial metallicity of the star because some $^{14}$N
%derives from conversion of the $^{12}$C produced inside the star and mixed
%up by the TDU. 
%
The other possible neutron source in AGB stars is the
$^{13}$C($\alpha$,n)$^{16}$O reaction. Formation of the required $^{13}$C
is usually obtained by allowing some extra mixing of protons at the deepest
extent of the convective envelope during each TDU episode
\citep{busso99}. However, we did not include the $^{13}$C pocket in our models 
because in IM-AGB stars the formation of the $^{13}$C pocket appears to be inhibited by p captures
occurring at the hot base of the convective envelope during the TDU
\citep{goriely04}. The surface $s$-process abundances increase with each
TDU episode as they are successively mixed into the convective envelope
(see Figure \ref{f:srf2}). The activation of the $^{22}$Ne neutron source
depends on the temperature at the base of the TP convection zone, which
increases with each TP. Because the $s$-process nuclei accumulate in the 
He intershell and are irradiated during each TP, they experience a total
neutron flux that increases with the TP number. As a result, the
$s$-process surface abundances show an increase initially of the elements
at the first $s$-process peak (Rb, Sr, Y, and Zr), then of the elements at
the second peak (e.g., Ba), and finally of those of the third peak at
Pb. 
The $s$-process elements belonging to the same neutron magic
peak (e.g., Sr, Y and Zr, N = 50)
are produced together along the $s$-process path, according to the local neutron capture rates.
In Figure \ref{f:srf2} we show the production of heavy 
elements for two 6 M$_{\odot}$ AGB models (standard case and $\alpha_{MLT}$ enanched+Bloecker mass-loss law).
For the first model, [Rb/Fe] increases up to $\sim$1.4 dex, Sr-Y-Zr up to $\sim$ 1.1 dex
[Ba/Fe] up to 0.7 dex and Pb only marginally changes, growing less than 0.2 dex.
In the second model, the $s$-process production is much weaker, 
since all elements under consideration have abundances lower than roughly 0.5 dex.  
To explain
variations in only one of these elements (eg., Y, as reported by VG11) would
require very unusual conditions. The production of Rb with
respect to solar is higher than the production of Y, due to the high
neutron densities resulting from the activation of the $^{22}$Ne neutron
source \citep{vanraai12}.

\subsubsection{Comparison with observations of M4}

In Figures \ref{f:yldnao5} to \ref{f:yldli6} we compare 
the elemental yields in the form of [X/Fe] predicted by                         
the present models with the key abundances observed in M4. 
The ``yields''
are defined as the integrated elemental abundance lost in the AGB winds
during the whole life of the star. This is the material that presumably
contributes to the gas from which the SG stars formed.

\begin{itemize}

\item[-] Na and O: In Figures \ref{f:yldnao5} and \ref{f:yldnao6} 
we present the comparison between
models and observations in the Na-O plane for the 5 M$_\odot$ and 6
M$_\odot$ models, respectively. It can be seen that the MLT5 model covers
the range of Na increase and O decrease, while the other 5 M$_\odot$ models
fail to deplete O sufficiently. On the other hand, most of the 6 M$_\odot$
models deplete enough O, but fail to produce significant amounts of Na. The
MBR model, computed by altering the p-capture reaction rates that affect
the Na abundance (within the uncertainties given by \citealt{iliadis10}),
results in a Na yield $\sim$ 70\% higher than using the recommended
rates. This is not enough to match the observed
variation between the FG and SG stars, however, it should be
considered as a conservative choice.  \citet{iliadis10} provided new
estimates for nuclear reaction rates based on a Monte Carlo model that
allowed the evaluation of the probability distribution function for each
rate. They defined as ``low'' and ``high'' those rates corresponding to a
coverage probability of 1$\sigma$ -- clearly the uncertainties would be larger
if one adopted a higher coverage probability. Also, it should be noted that
while the Monte Carlo method takes into account all current experimental
information, it cannot be excluded that different information may arise
from future experiments, particularly for the
$^{22}$Ne(p,$\gamma$)$^{23}$Na and $^{23}$Na(p,$\alpha$)$^{24}$Mg rates,
which are affected by low-energy resonances that are difficult to reach
experimentally.

\begin{figure*}
\centering
    \begin{minipage}[t]{0.45\linewidth}
      %\centering
       \includegraphics[width=0.8\textwidth]{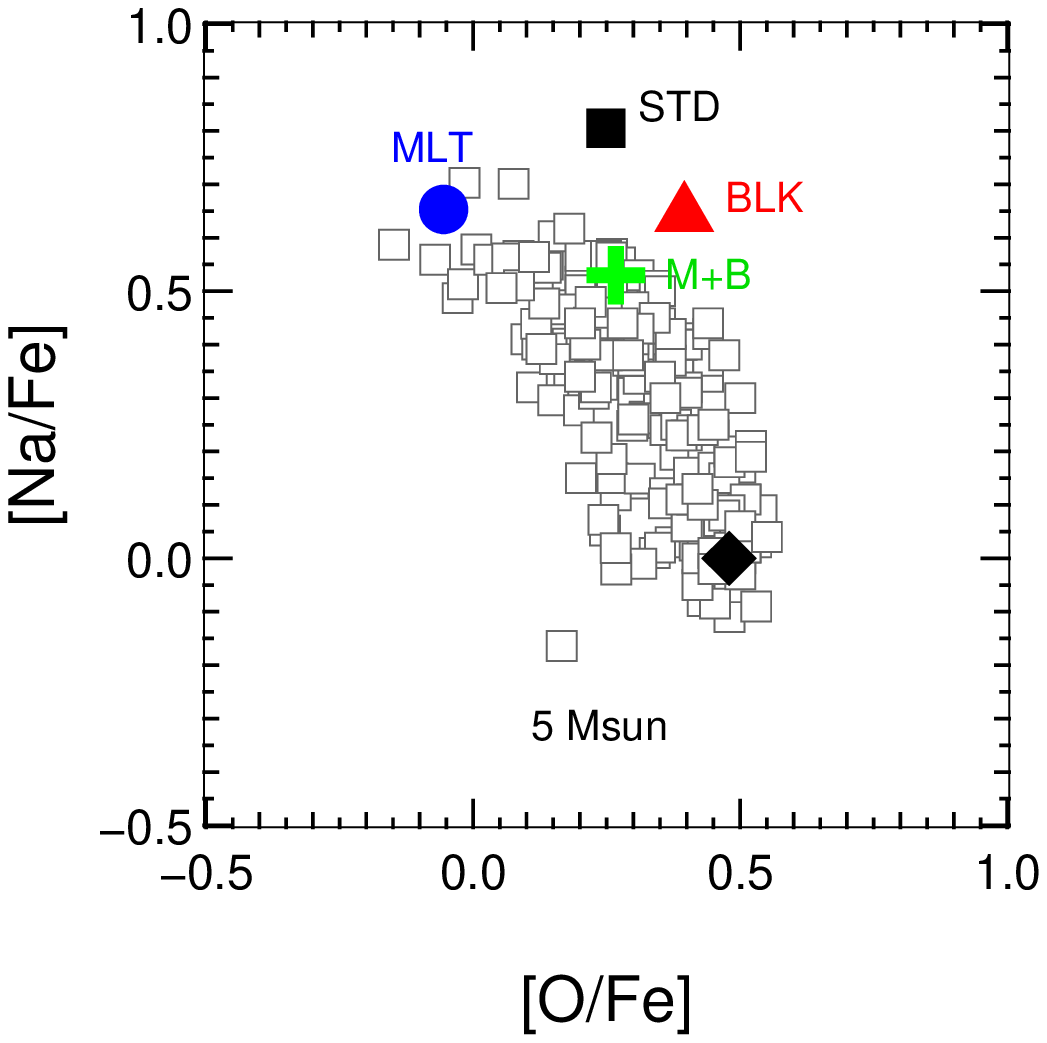}
       \caption{Comparison between our 5 M$_\odot$ model yields and observations
  in the O-Na plane (as in Figure~\ref{f:nao}). Large markers indicate yields from models as indicated
  in the figure (see Table \ref{t:models}), except for the solid diamond
  which shows the initial composition. Observational data are from the same
  sources as in Figure \ref{f:nao}.}\label{f:yldnao5}
    \end{minipage}
    \begin{minipage}[t]{0.45\linewidth}
      %\centering
      \includegraphics[width=0.8\textwidth]{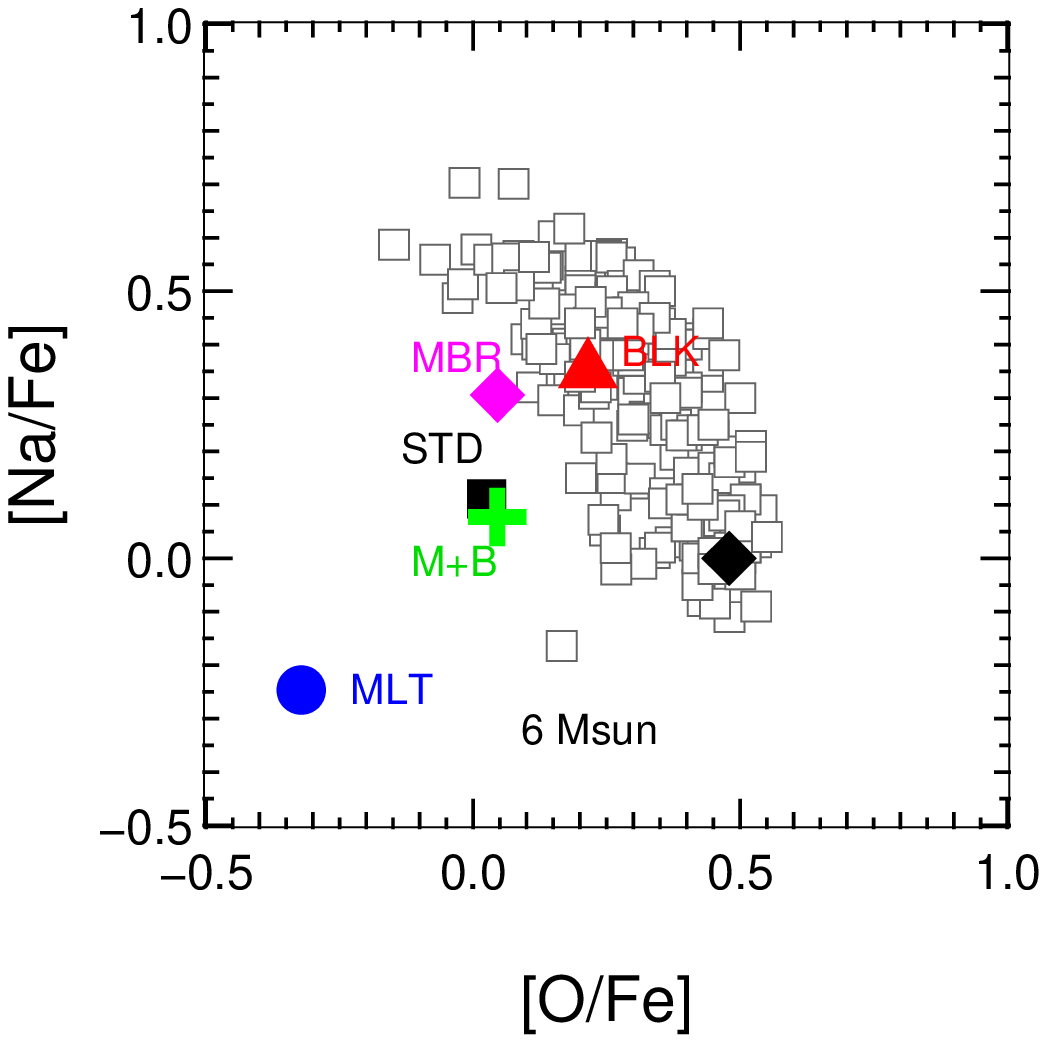}
   \caption{Same as Figure \ref{f:yldnao5} except for the 6 M$_\odot$ models.}\label{f:yldnao6}
    \end{minipage}
\end{figure*}

\begin{figure*}
\centering
    \begin{minipage}[t]{0.45\linewidth}
      \centering
       \includegraphics[width=0.8\textwidth]{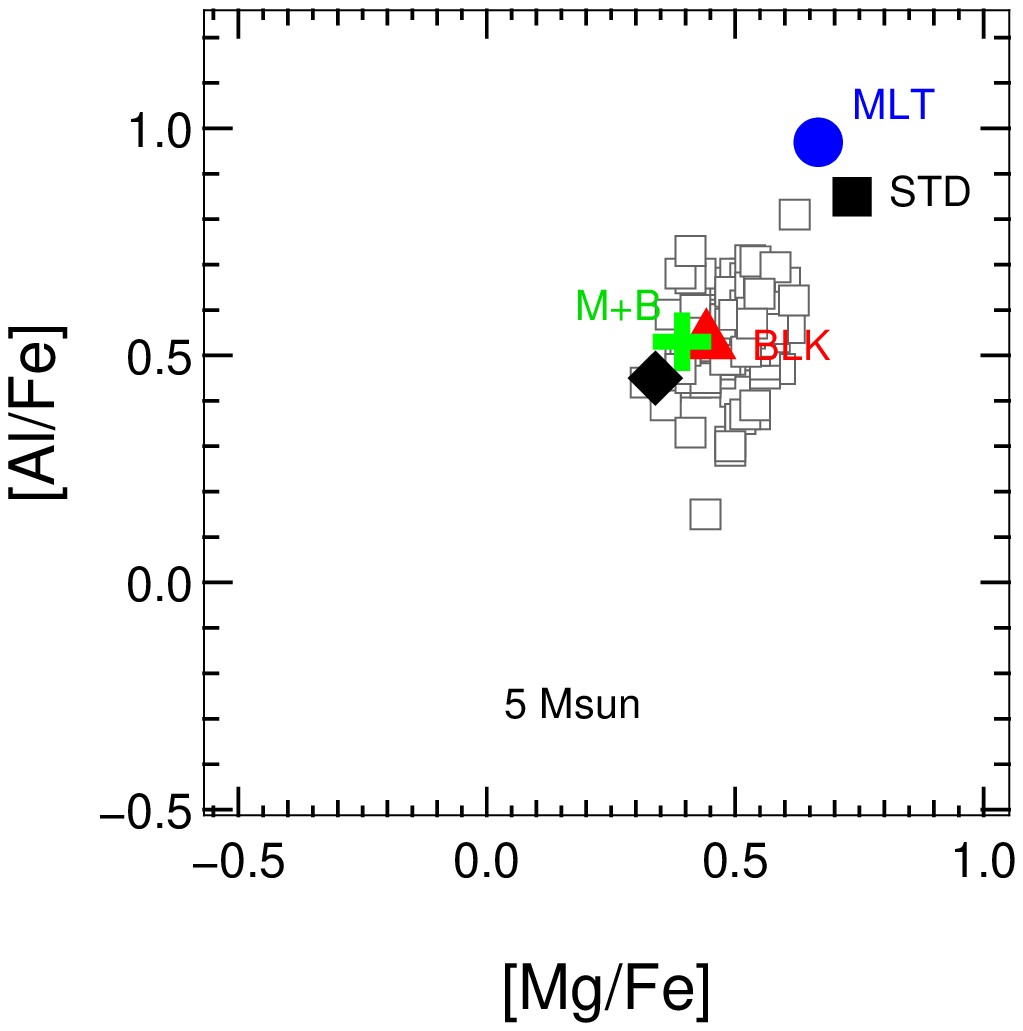}
       \caption{Same as Figure \ref{f:yldnao5} but for the Mg-Al plane. Note that we have offset the initial Al composition 
               used in the models to the value observed in 
the FG stars of M4.}\label{f:yldmgal5}
    \end{minipage}
    \begin{minipage}[t]{0.45\linewidth}
      \centering
      \includegraphics[width=0.8\textwidth]{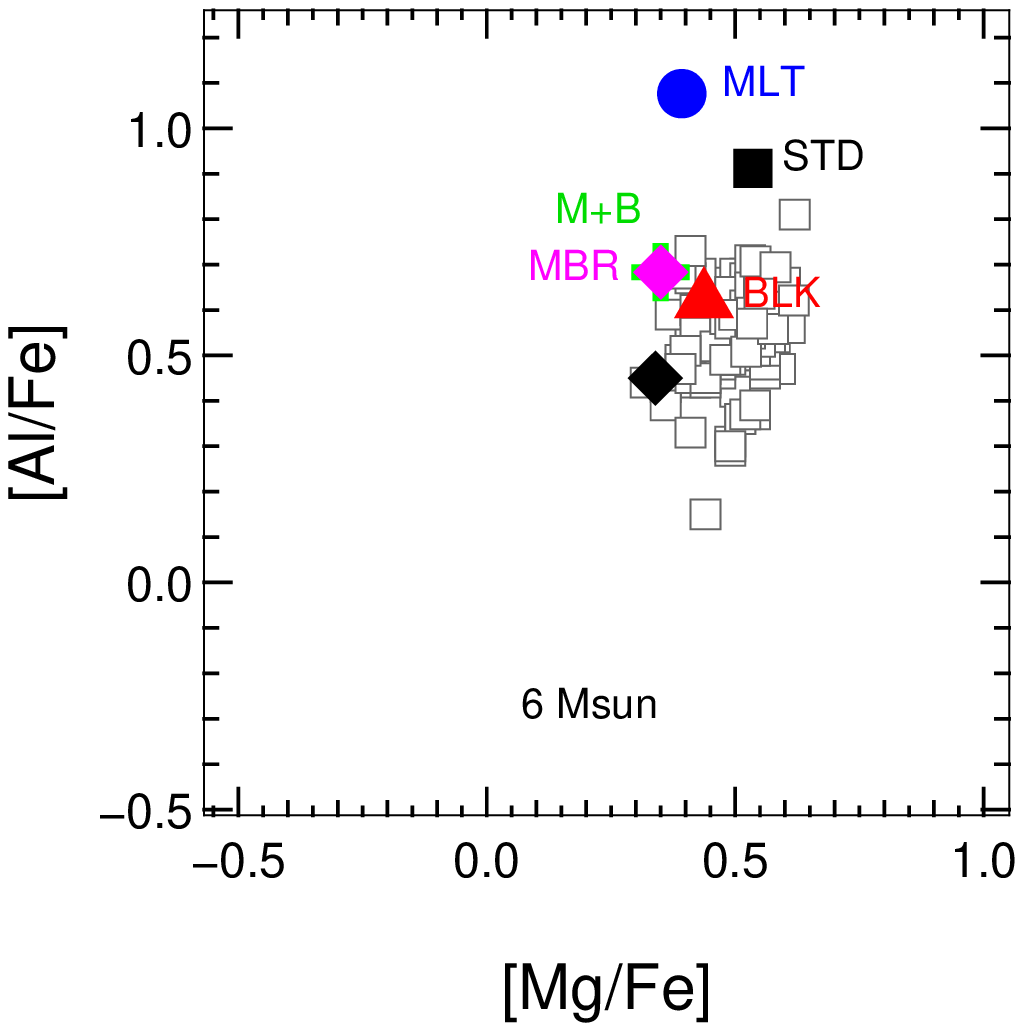}
      \caption{Same as Figure \ref{f:yldmgal5} but for the 6 M$_{\odot}$ models.}\label{f:yldmgal6}
    \end{minipage}
\end{figure*}

\item[-] Mg and Al: In Figures \ref{f:yldmgal5} and \ref{f:yldmgal6} we present the comparison
between models and observations in the Mg-Al plane for the 5 M$_\odot$ and
6 M$_\odot$ models, respectively. The STD and MLT models produce Mg and/or
Al abundances too high with respect to the observations, either due to
dredge-up of $^{25,26}$Mg produced by $\alpha$ captures on $^{22}$Ne, in
the case of the 5 M$_\odot$ models, or due to efficient HBB in case of the
6 M$_\odot$ models. 
On the other hand, all the BLK and M+B models are 
well within the observations (we remind the reader that the extent of
the Al spread is debated, as described in detail in
Section~\ref{sec:light}).
\begin{figure*}
\centering
    \begin{minipage}[t]{0.45\linewidth}
      \centering
       \includegraphics[width=0.8\textwidth]{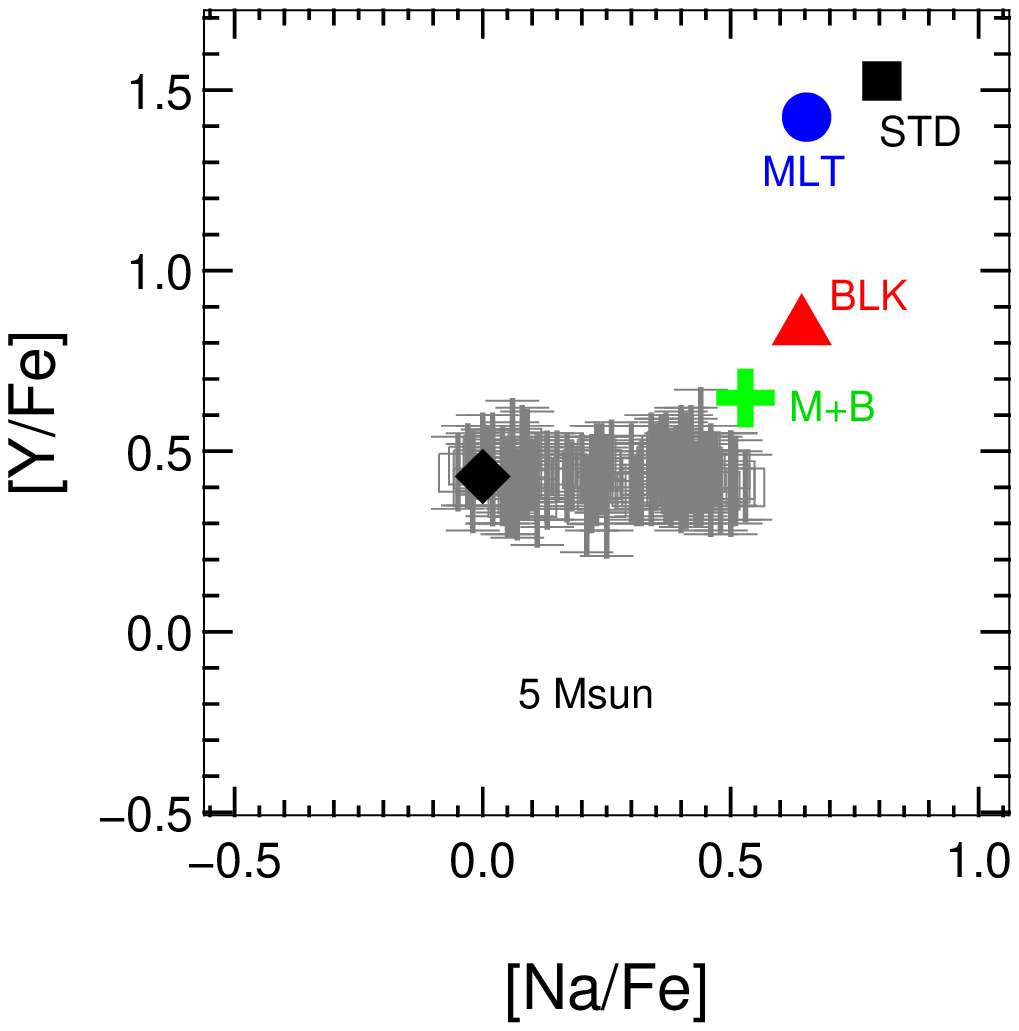}
       \caption{Same as \ref{f:yldnao5} but for the Na-Y plane. 
Observational data are from the current study for Y and from Ma08 for Na. Note that we have offset the initial Y composition used in the models to the value observed in the FG stars of M4.}\label{f:yldyna5}
    \end{minipage}
    \begin{minipage}[t]{0.45\linewidth}
      \centering
      \includegraphics[width=0.8\textwidth]{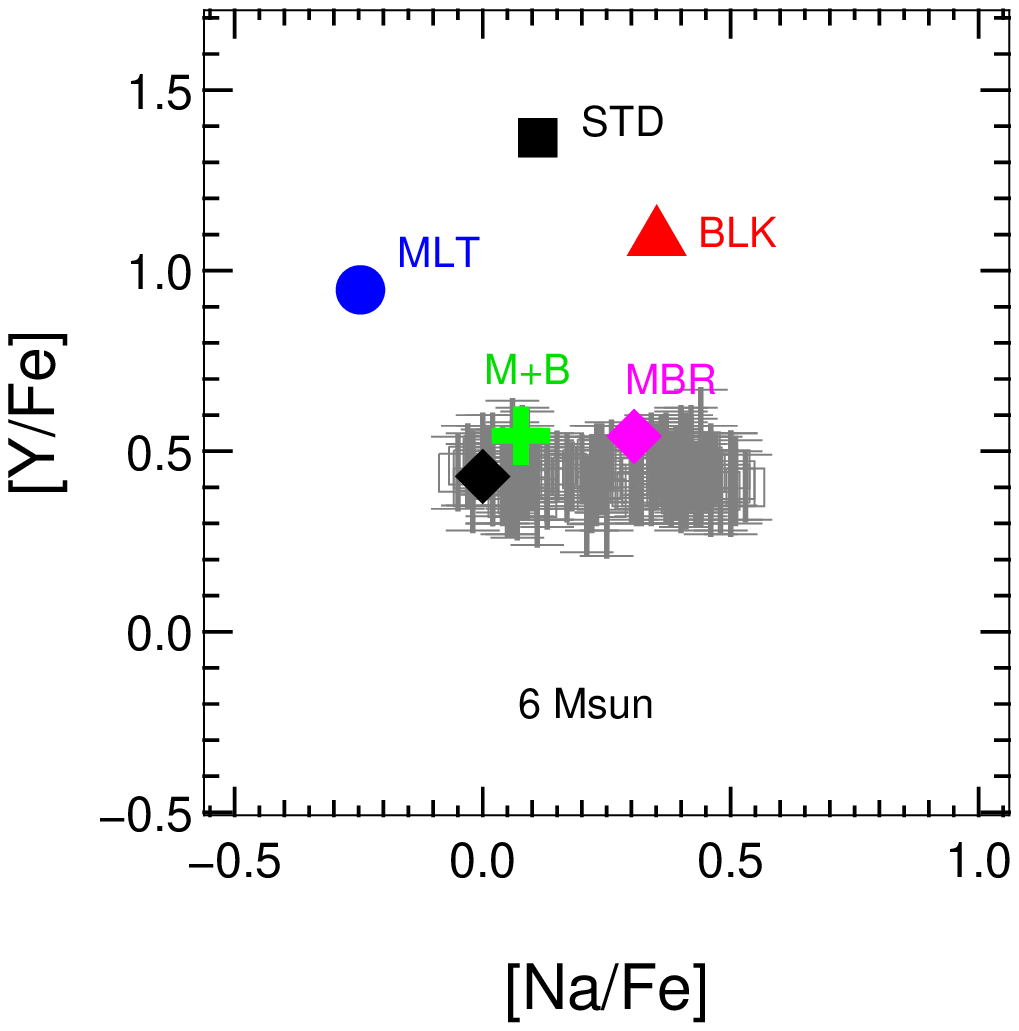}
      \caption{Same as Figure \ref{f:yldyna5} but for the 6 M$_\odot$ models}\label{f:yldyna6}
    \end{minipage}
\end{figure*}
\item[-] Na and Y: In Figures~\ref{f:yldyna5} and \ref{f:yldyna6} present the comparison between
models and observations in the Na-Y plane for the 5 M$_\odot$ and 6
M$_\odot$ models. The STD and MLT models, which experience a
large number of TDU episodes, produce Y variations an order of magnitude
higher than the spread observed in M4. Therefore, within the IM-AGB pollution scenario these models would cause 
a larger s-process enrichment in SG stars compared to FG stars, which is not
observed according also to our present results.
On the other hand, the BLK and M+B
models, which experience fewer TDU episodes, produce Y variations of $< 0.3$
and 0.2 dex for the 5 M$_\odot$ and 6 M$_\odot$ models respectively. These models may reproduce better the observations,
i.e., the missing extra-enrichment in $s$-process elements 
of SG stars compared to FG stars in M4 within the uncertainties.
A similar conclusion can be derived
using the Na-Ba plane (Figures \ref{f:yldbana5} and \ref{f:yldbana6}).
\begin{figure*}
\centering
    \begin{minipage}[t]{0.45\linewidth}
      \centering
       \includegraphics[width=0.8\textwidth]{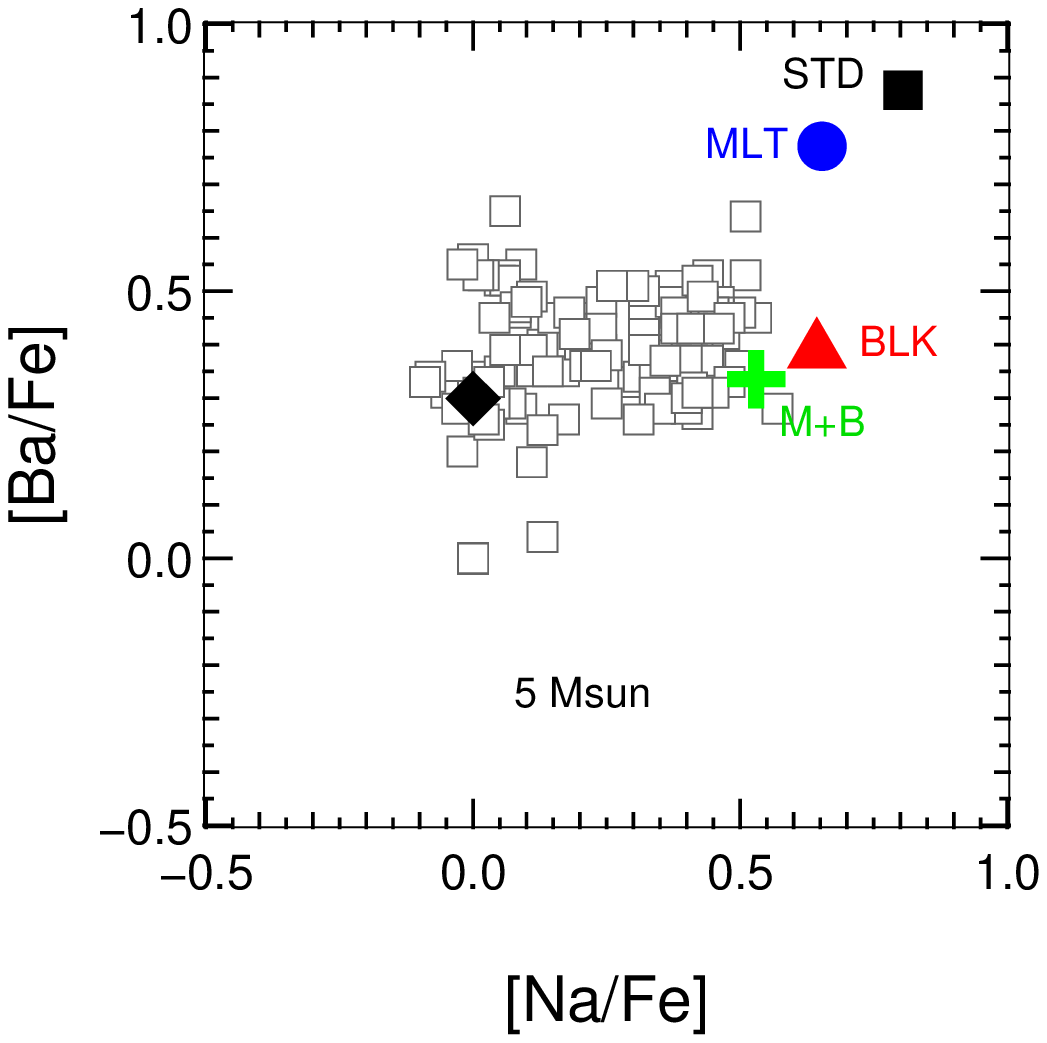}
       \caption{Same as Figure \ref{f:yldnao5} but for the Na-Ba plane. 
Observational data are the same as in Figure \ref{f:BaFe}. Note that we have offset the initial Ba composition used in the models to the value observed in the FG stars of M4.}\label{f:yldbana5}
    \end{minipage}
    \begin{minipage}[t]{0.45\linewidth}
      \centering
      \includegraphics[width=0.8\textwidth]{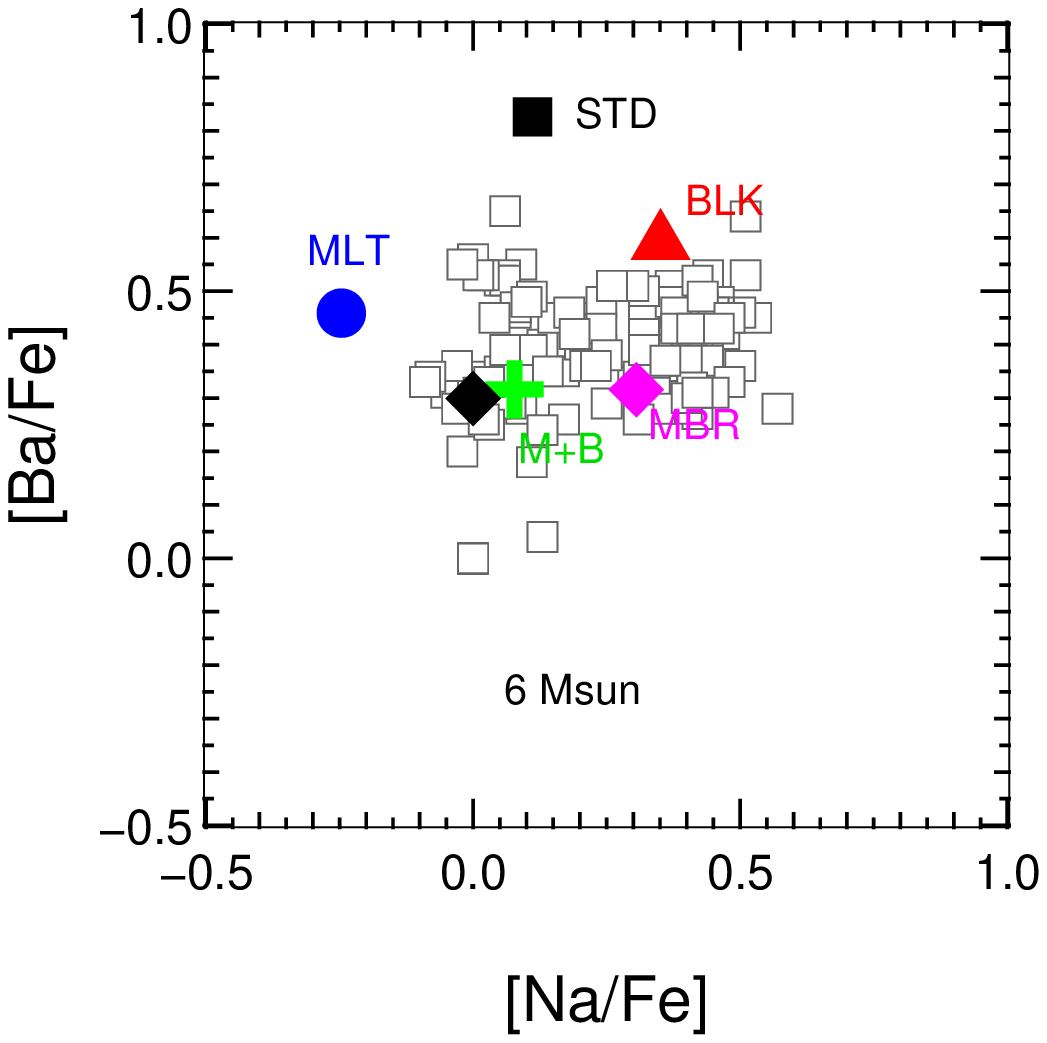}
      \caption{Same as Figure \ref{f:yldbana5} but for the 6 M$_\odot$ models.}\label{f:yldbana6}
    \end{minipage}
\end{figure*}
\begin{figure*}
\centering
    \begin{minipage}[t]{0.45\linewidth}
      \centering
       \includegraphics[width=0.8\textwidth]{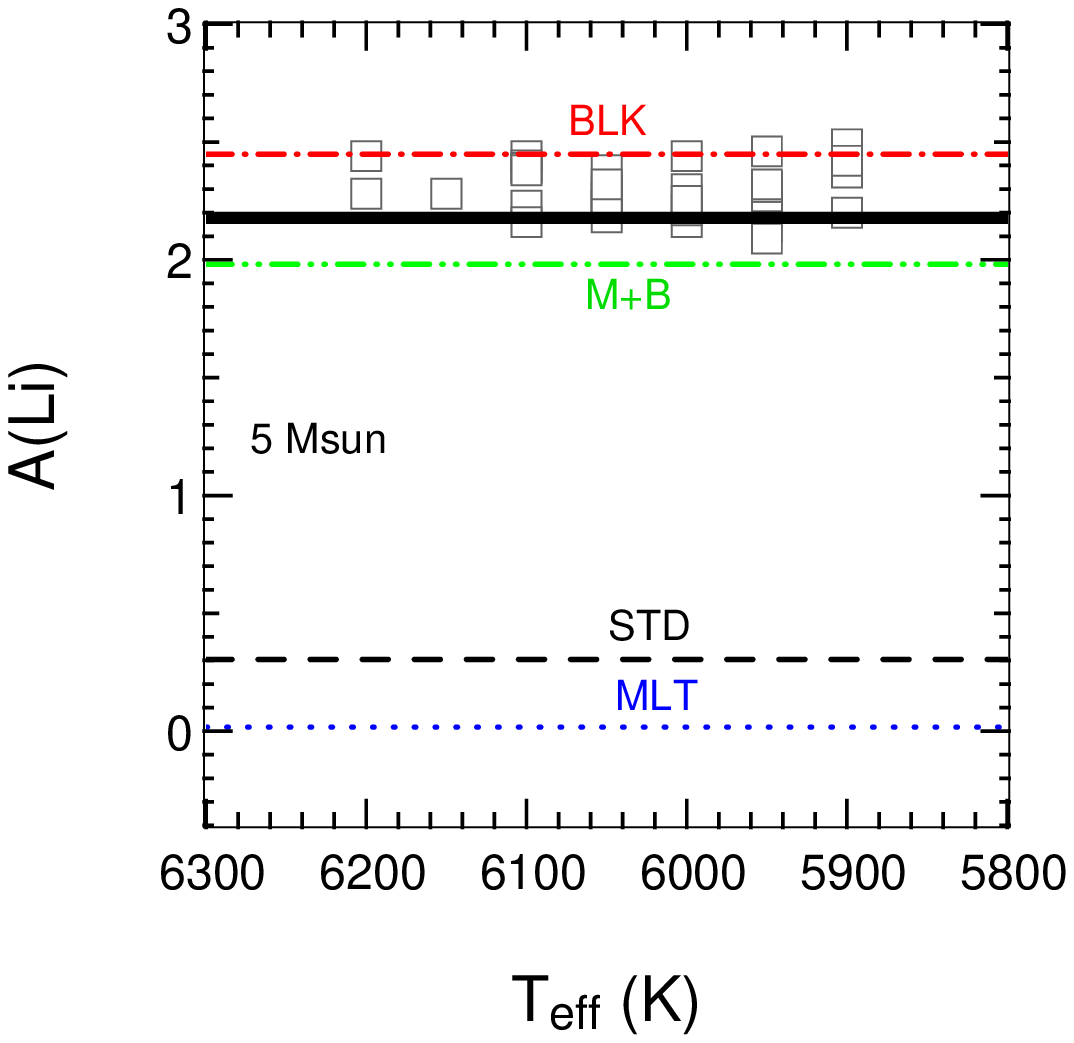}
       \caption{Same as Figure~\ref{f:yldnao5} but for the T$_{eff}$-Li plane. 
The thick, solid line is the initial abundance of Li, the other lines
represent the predicted yields as indicated by the labels.
Observational data are the dwarf stars discussed in Section~\ref{sec:light}.}\label{f:yldli5}
    \end{minipage}
    \begin{minipage}[t]{0.45\linewidth}
      \centering
      \includegraphics[width=0.8\textwidth]{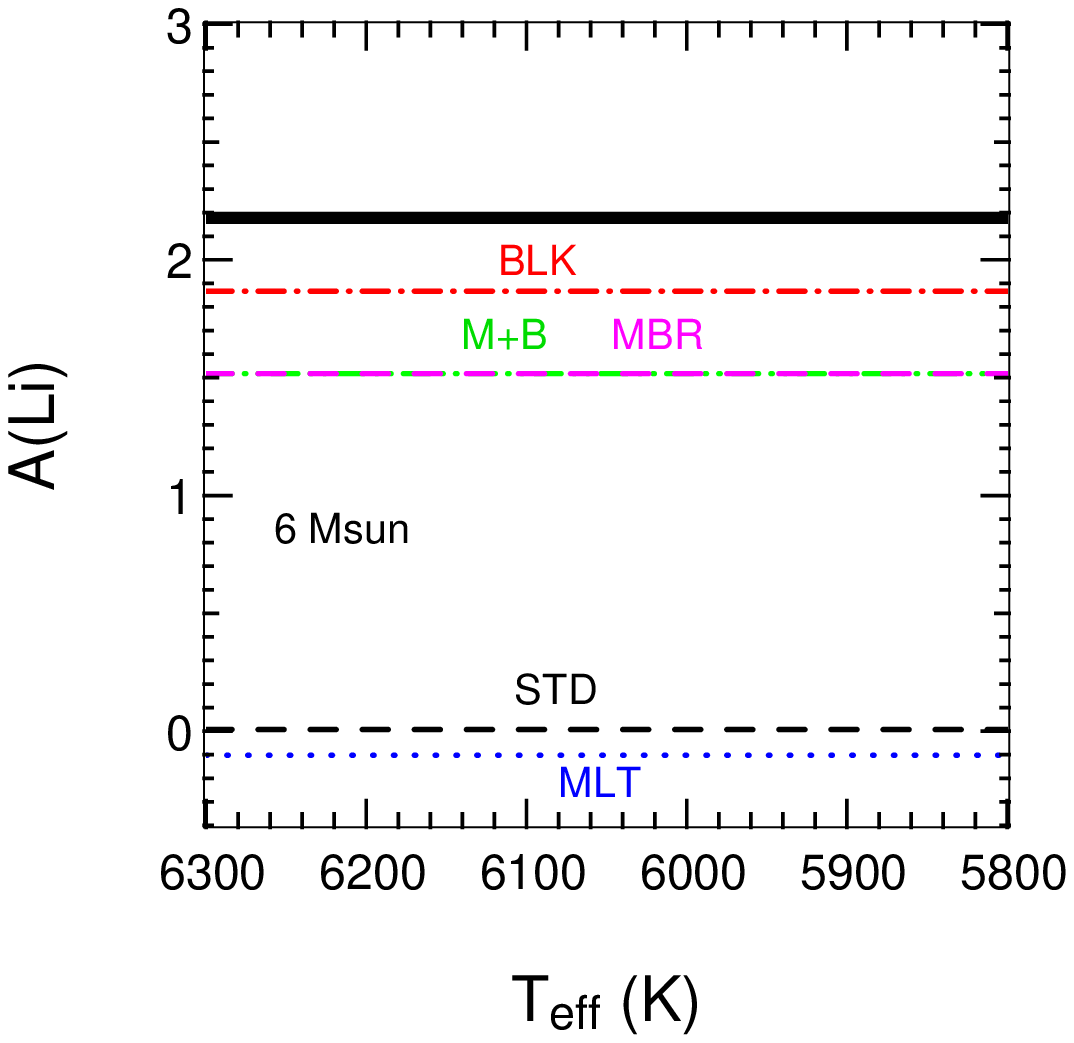}
      \caption{Same as Fig \ref{f:yldli5} for the 6 M$_\odot$
  models.}\label{f:yldli6}
    \end{minipage}
\end{figure*}
\item[-] Li: Figures~\ref{f:yldli5} and \ref{f:yldli6} show the comparison between
models and observations in the T$_{eff}$-Li plane for the 5 M$_\odot$ and 6
M$_\odot$ models. All our models present a brief phase of Li enrichment at
the stellar surface, during the first few TPs, with a peak of A(Li)$\sim$4 dex. 
The final Li yield depends
on when the star loses most of its mass
(\citealt{travaglio01}; \citealt{ventura10}). The STD and MLT models, which experience extended
lifetimes, result in yields close to zero, because they lose most of their
mass after Li has been destroyed by HBB. Conversely, the BLK and M+B
models, which have shorter lifetimes, eject a substantial amount of mass
during their Li-rich phases, producing relatively high yields -- close to
the initial value -- particularly when considering the 5 M$_\odot$ model.
\end{itemize}

Overall, none of our STD and MLT models can reproduce the observed
constancy in the $s$-process and Li abundances. Furthermore, they produce
Mg and Al outside the observed variations. 
%
%
%If we strongly reduced the
%efficiency of the TDU episodes (another model uncertainty we did not
%investigate here) the discrepancy with the $s$-process constraint could be
%resolved, however, the problem of the Li destruction would remain. 
%
%
On the other hand, the BLK and M+B models can reproduce most of the observational
constraints, although the 5 M$_\odot$ models do not deplete enough O and
result in slightly too high $s$-process abundances, while the 6 M$_\odot$
models require some adjustment in the reaction rates to explain the Na
spread, and they deplete too much Li. The Li problem might be rectified by
a further slight increase in mass-loss rate, for instance using a slightly
higher value of $\eta$ in the Bloecker formula, so that more mass is lost
during the Li-rich phase, and/or by considering in detail the uncertainties
in the nuclear reaction rates involved.

Finally, we report that in the M+B6 model the [Rb/Fe] increases 
of $\sim$ +0.2 dex, slightly higher than that predicted for
[Y/Fe]. 
However, this is not a problem since such a weak overall enrichment of
Rb and Y in these IM-AGB models as polluters would likely not be
observable in SG stars.
For the same model, the change in the sum of C+N+O is $\sim$ +0.3
dex, and the He enrichment is $\delta$Y $ = +0.09$. 
All the models produce [N/Fe] $\simeq$1.5-2 
(the observed variation is up to $\simeq$1 dex; e.g., VG11), but none
of them deplete C to the observed level of 
[C/Fe]$\simeq$$-$0.5 (e.g., VG11). 
The lowest [C/Fe] obtained is $\simeq$ 0 using the M+B6 model.

\subsubsection{Comparison with other observational constraints}

To ascertain if our IM-AGB models computed including efficient
HBB and strong mass loss are a realistic solution for the anomalies
observed in M4 we can, in principle, check if these models are consistent
with other observational constraints. A problem however is that there are
very few direct observations of abundances in IM-AGB stars, and these are for stars in the Galaxy, LMC, or SMC, i.e., of metallicities higher than M4. 
One study that
we are aware of is that of \cite{mcsaveney07}, who obtained spectra of a
sample of LMC and SMC AGB stars. The severe difficulties in performing
these observations are clear from the \cite{mcsaveney07} study, where they
attempted to measure abundances in the cool AGB atmospheres (which are also
dynamic/pulsating, see e.g., \citealt{lebzelter10}) of 7 stars but were only
able to obtain results for two. The results from their two stars support an increase in $\alpha_{MLT}$ at
least as high as we have used in our MLT models and also the phenomena
of HBB and TDU as seen in the models. Other studies of IM-AGB stars are those of
\citet{plez93} for the Small Magellanic Cloud (SMC), and those of
\citeauthor{garcia06} (2006, 2007, 2009)\nocite{garcia07}\nocite{garcia09}
discussed below, for the Galaxy, the SMC and the Large Magellanic Cloud (LMC). 
More accessible are planetary nebulae (PNe) of Type I. 
The material surrounding the compact
central stars in these objects is believed to consist of the material 
ejected by IM-AGB stars. They are also located in the Galactic disk
\citep{stanghellini06,sterling08}, or in the LMC and SMC~\citep{kaler90}. 
Again, all these objects have
metallicities higher than M4, which may affect the uncertain parameters
related to the evolution of these stars.  Keeping this caveat in mind, we
can examine the available constraints.

Planetary nebulae (PNe) of Type I are believed to be IM stars that have
completed the AGB phase, have little or no stellar envelope left, and are
evolving blue-wards at constant luminosity in the CMD. PNe of Type I show the
clear signature of HBB in their He/H and N/O ratios (e.g.,
\citealt{kaler90,stanghellini06}) and no enhancements of
the $s$-process elements Se and Kr, which are close to the first
$s$-process peak.  This is well explained by models of IM-AGB stars of
solar metallicity \citep{karakas09}. However, this simple picture has been
somewhat complicated by the detection of extremely high abundances of Rb in
OH/IR stars in the Galaxy \citep{garcia06}, and in the SMC/LMC \citep{garcia09}, which represents the first proof of
the activation of the $^{22}$Ne neutron source in IM-AGB stars. The IM-AGB
nature of these stars was determined on the basis of the velocities of their
OH masers, their location in the Galactic plane (as indicative of belonging
to a young population), and the presence of Li produced by HBB, at least in
some of the objects. The very high Rb overabundances, with [Rb/Fe] up to
+2.5 dex for Galactic stars and up to +5 dex for stars in the Magellanic
Clouds are however not matched by standard IM-AGB models
\citep{garcia09,vanraai12}. \citet{karakas12} showed that if the stellar
lifetime is extended by lowering the mass-loss rate, so that a larger
number of TDUs is allowed, and the rate of the
$^{22}$Ne($\alpha$,n)$^{25}$Mg reaction is taken from the NACRE compilation
\citep{angulo99}, then it is possible to reach [Rb/Fe] $\simeq 1.4$, close to
the average [Rb/Fe] observed in the Galactic OH/IR stars.  Clearly, we 
have here a contradictory situation since to match these observations the
$s$-process nucleosynthesis in IM-AGB stars needs to be enhanced, while to
match the constancy of the $s$-process elements in M4, including Rb, the
$s$-process nucleosynthesis in IM-AGB stars needs to be suppressed.

We note however that there are still several problems related to the
interpretation of the high Rb abundances in IM-AGB:

\begin{enumerate}

\item{as discussed in detail by \citet{garcia09} and \citet{vanraai12}, 
  serious problems are present in current model atmospheres of luminous AGB
  stars since these models are performed in 1D and do not include important
  effects such as the presence of a circumstellar dust envelope and dust
  formation, which may lead to systematic uncertainties.  More realistic
  model atmospheres for intermediate-mass O-rich AGB stars (e.g., the
  inclusion of a circumstellar dust envelope and 3D hydrodynamical
  simulations, see e.g., \citealt{lebzelter10}) as well as NLTE calculations
  need to be developed;}

\item{the IM-AGB models that predict high Rb enhancements also necessarily
  predict some Zr enhancements, which are not observed. 
  These stars show [Zr/Fe] $<0.5$ \citep{garcia07};}

\item{the NACRE $^{22}$Ne($\alpha$,n)$^{25}$Mg rate is probably too high
  when considering the more recent experiments \citep{jaeger01};}

\item{Li abundances in the observed stars present a large range, from
  depletion to enhancements, however (as also discussed above in relation
  to the present models), Li enhancements can only be seen early during the
  AGB evolution, while high Rb enhancements are obtained only after many
  TPs.}

\end{enumerate}

Another possibility is that the TDU
efficiency and the mass-loss rate are very sensitive to the initial stellar
mass and the metallicity. For example, more massive IM-AGBs than those
considered here and Super-AGB stars (\citealt{pumo08}, \citealt{doherty10}, 
\citealt{siess10}) might experience less
efficient TDU and/or higher mass-loss rates and be responsible for the
pollution of GCs. Obviously, more models of IM-AGB stars and observations
of both Rb-rich stars and Type I PNe will help us to understand these
issues.  For the time being it seems difficult to use the high-Rb
observations as a strong constraint to extrapolate information to IM-AGB
stars in GCs and viceversa.

A further difficulty when comparing to other observational constraints is
the possibility that IM-AGB stars could evolve differently in clusters and
in the field. It is known that B stars (which eventually evolve to be
IM-AGB stars) in open clusters typically rotate more rapidly than B stars
in the field (though this may be due to their being in a less evolved phase of
their evolution, e.g., \citealt{huang10}). Fast rotation would have a strong effect
on the stellar evolution prior to the AGB phase possibly producing too high
C+N+O yields to be compatible with the observations, as shown by
\citet{decressin09}, though the effect of magnetic fields was not included
in these models. Moreover, binary interaction is important in shaping stellar yields (\citealt{izzard06}, \citealt{vanbeveren12}) and its effects would be different if the binary distribution was different for field and cluster stars.

\section{Summary and concluding remarks}\label{sec:summary}

In summary we have considered four observational constraints and 
derived the following results for M4:

\begin{enumerate}
   \item{In the O-Na plane, the MLT5 model provides a good fit to the
     observed spread. Also the STD6 and the M+B6 models may result in a
     good fit, provided that the rates of the nuclear reaction that produce
     and destroy Na during HBB are varied somewhat beyond the `high' and
     `low' values given by \cite{iliadis10}. These variations are within
     current possibilities and need to be investigated.}

   \item{In the Mg-Al plane, all the STD and MLT models produce variations
     in Mg and/or Al outside the observed spread, while all the BLK and M+B
     models are within the observations.}

   \item{As in item (ii), in the Na-Y and Na-Ba planes the STD and MLT
     models produce variations outside the observed spread, while the BLK
     and M+B models can explain the constancy of $s$-process abundances
     between generations, particularly when considering the 6 M$_\odot$
     models.}

   \item{As in items (ii) and (iii), when considering Li the STD and MLT
     models result in no Li production, while the BLK and M+B models can
     explain the constancy of the Li abundance between generations,
     particularly when considering the 5 M$_\odot$ models.}
     
   \item{All the models result in production of He beyond the observed variations, which indicates the need of some dilution with primitive material. The change in the sum of C+N+O may be comparable to the observations, within the uncertainties, only if we consider the BLK and M+B models. 
The observed depletion of C by a factor $\sim$ 2 is not obtained by any of the models.}  

\end{enumerate}

Although none of these models can simultaneously match all 
the constraints listed above, we believe that our
exploration of the model uncertainties shows that IM-AGB stars cannot be ruled 
out as potential polluters for the M4 SG population since reasonable 
variations in input physics can still provide a reasonable agreement between the
theoretical yields and observations. 
More work is clearly needed to improve the input physics of IM-AGB stars.
We note, however, that the fact that SG stars probably formed from the ejecta of a range of masses (not a single mass) and that there must be a certain dilution between pristine and polluted material (see e.g., \citealt{gratton12}), do not allow us to simultaneously reproduce all the observational constraints any better with the present models. 

We have confirmed the
previous finding by \citeauthor{ventura05c} (\citeyear{ventura05c}; 2010) that IM-AGB
models computed including efficient HBB and strong mass loss (e.g., the
Bloecker mass-loss rate) can provide a match to the observations of the
light elements. Furthermore, we have shown that these same models can
also be consistent with the constraint that there are no
variations in the $s$-process elements in M4 between FG and SG. 
When comparing to
observational constraints other than GCs, however, it appears difficult to
establish a self-consistent scenario that could explain all the current
observations related to IM-AGB stars.  Also the question remains of 
why IM-AGB stars in GCs should experience a stronger mass loss than 
what is observed in the Galaxy and in the Magellanic Clouds \citep{vw93}. 
In any case, pursuing such
comparison offers us the future opportunity of better understanding all of
the current issues related to nucleosynthesis in IM-AGB stars via more
refined models and observations.

In particular, AGB $s$-process models similar to those we have explored
here are needed for a more extended range of masses, i.e., between 3
M$_\odot$ and 9 M$_\odot$, and for different choices of the mass-loss rate
and efficiency of convection. The models of \cite{lugaro12} have shown that
as the stellar mass decreases the effect
of the TDU (producing C, F, and $s$-process elements) gradually becomes
predominant over the effect of HBB (depleting O, and F, and producing Li).
AGB models of slightly lower mass than those presented here have been
invoked as a possible explanation for variations in the $s$-process
elements in M22 (\citealt{roederer11}, \citealt{marino09}), as well as for their correlation with
fluorine (\citealt{dorazi13}). A large grid of AGB models 
and realistic models of the gas dynamics in the forming GC 
(\citealt{krause12},\citeyear{krause13}) will be fundamental
to ascertain at which mass we can see the raising of the $s$-process and
fluorine abundances, and to derive information on the timescales of
formation of the different populations in GCs.

\section*{Acknowledgments}
The authors warmly thank the referee for a very careful reading of the manuscript and for valuable comments and
suggestions.
This work made extensive use of the SIMBAD, Vizier, and NASA ADS database.
We acknowledge B. Plez for providing his unpublished line lists.  We thank
F. D'Antona, T. Decressin, C.  Doherty, A. Garc{\'{\i}}a-Hern{\'a}ndez, A. I. Karakas, G. Imbriani, 
A.F. Marino, M. Wiescher for helpful discussions and C. Iliadis for detailed explanation of reaction rate
uncertainties.  ML is supported by an ARC Future Fellowship and a Monash
Fellowship. SWC is supported by an Australian Research Council Discovery
Project grant (DP1095368). MP thanks support from an Ambizione grant of the SNSF (Switzerland), and from EuroGENESIS (MASCHE).

%\clearpage
%\begin{thebibliography}{99}

\bibliographystyle{mn2e} % style mn2e.bst
\bibliography{draft_rev4} % references in refsM4.bib

\label{lastpage}

\end{document}